\documentclass[aps,pra,twocolumn,superscriptaddress,longbibliography,floatfix]{revtex4-1}

\usepackage{amsmath,amssymb,graphicx,bm,dsfont}
\usepackage{xcolor}
\usepackage[colorlinks=true,urlcolor=blue,linkcolor=blue,citecolor=blue,bookmarks=false]{hyperref}

\usepackage[english]{babel}

\usepackage{natbib}
\usepackage{ dsfont }
\usepackage{textcomp}

\begin{document}
	
\title{
Field-induced reorientation of helimagnetic order in Cu\textsubscript{2}OSeO\textsubscript{3} probed by magnetic force microscopy
}
	
	\author{Peter~Milde}
	\email{peter.milde@tu-dresden.de}
	\affiliation{Institute of Applied Physics, Technische Universit\"at Dresden, D-01062 Dresden, Germany}
	
	\author{Laura~K\"{o}hler}
	\affiliation{Institute of Theoretical Physics, Technische Universit\"{a}t Dresden, D-01062 Dresden, Germany}
	\affiliation{Institute for Theoretical Solid State Physics, Karlsruhe Institute of Technology, D-76131 Karlsruhe, Germany }

	\author{Erik~Neuber}
	\author{P.~Ritzinger}
	\affiliation{Institute of Applied Physics, Technische Universit\"at Dresden, D-01062 Dresden, Germany}
	
	\author{Markus~Garst}
	\affiliation{Institute of Theoretical Physics, Technische Universit\"{a}t Dresden, D-01062 Dresden, Germany}
	\affiliation{Institute for Theoretical Solid State Physics, Karlsruhe Institute of Technology, D-76131 Karlsruhe, Germany }
	\affiliation{Institute for Quantum Materials and Technology, Karlsruhe Institute of Technology, D-76344 Eggenstein-Leopoldshafen, Germany}
	
	\author{Andreas~Bauer}
	\author{Christian~Pfleiderer}
	\affiliation{Physik-Department, Technische Universit\"{a}t M\"{u}nchen, D-85748 Garching, Germany}
	
	\author{H.~Berger}
	\affiliation{Institut de Physique de la Mati\`ere Complexe, \'Ecole Polytechnique F\'ed\'erale de Lausanne, 1015 Lausanne, Switzerland}

	\author{Lukas~M.~Eng}
	\affiliation{Institute of Applied Physics, Technische Universit\"at Dresden, D-01062 Dresden, Germany}
	\affiliation{Dresden-W\"{u}rzburg Cluster of Excellence -- Complexity and Topology in Quantum Matter (ct.qmat), TU Dresden, 01062 Dresden, Germany}
	
\date{\today}
	
\begin{abstract}
Cu\textsubscript{2}OSeO\textsubscript{3} is an insulating skyrmion-host material with a magnetoelectric coupling 
giving rise to an electric polarization with a characteristic dependence on the magnetic field $\vec H$.
We report magnetic force microscopy imaging of the helical real-space spin structure on the surface of a bulk single crystal of CuO\textsubscript{2}SeO\textsubscript{3}. In the presence of a magnetic field, the helimagnetic order in general reorients and acquires a homogeneous component of the magnetization, resulting in a conical arrangement at larger fields. We investigate this reorientation process at a temperature of 10~K for fields close to the crystallographic $\langle 110\rangle$ direction that involves a phase transition at $H_{c1}$. Experimental evidence is presented for the  formation of magnetic domains in real space as well as for the microscopic origin of relaxation events that accompany the reorientation process.  
In addition, the electric polarization is measured by means of Kelvin-probe force microscopy. We show that the characteristic field dependency of the electric polarization originates in this helimagnetic reorientation process. Our experimental results are well described by an effective Landau theory previously invoked for MnSi, that captures the competition between magnetocrystalline anisotropies and Zeeman energy. 
\end{abstract}
	
	
\maketitle

\section{Introduction}

In the limit of weak spin-orbit coupling $\lambda_{\rm SOC}$ the cubic chiral magnets like MnSi \cite{Muehlbauer2009}, Fe\textsubscript{1-x}Co\textsubscript{x}Si \cite{Grigoriev2007,Muenzer2010}, FeGe \cite{Lebech_1989,Yu2010}, and Cu\textsubscript{2}OSeO\textsubscript{3} \cite{Seki2012a,Adams2012} are  dominated by only two coupling constants, the symmetric and antisymmetric exchange interaction, $J$ and $D$, respectively \cite{BauerPfleiderer2016}. Whereas the symmetric exchange is of zeroth order in $\lambda_{\rm SOC}$, the Dzyaloshinskii-Moriya interaction is of first order $D \sim \mathcal{O}(\lambda_{\rm SOC})$. Their ratio determines the characteristic wavevector $Q = D/J$ of the helimagnetic order that develops at zero magnetic field $\vec H = 0$. For finite $\vec H$, the helix assumes a conical arrangement until it is fully polarized at the internal critical field $\mu_0 H^{\rm int}_{c2} \simeq \frac{D^2}{J M_s}$, with $M_s$ the saturation magnetization. In addition, a small pocket of the skyrmion lattice phase just below the ordering temperature $T_{c}$ is realized at intermediate values of $\vec H$.

As a result, the above-mentioned materials share a very similar magnetic phase diagram. Details of it, however, depend on corrections that are parametrically smaller in $\lambda_{\rm SOC}$. In particular, the orientation of the helimagnetic order at zero field is determined by magnetocrystalline anisotropies, that are at least of fourth order in $\lambda_{\rm SOC}$, and generally favour the spin spiral to align either along a crystallographic $\langle 111\rangle$ or $\langle 100 \rangle$ direction like, e.g., in MnSi or Cu\textsubscript{2}OSeO\textsubscript{3}, respectively. The Zeeman energy competes with the magnetocrystalline anisotropies resulting in a reorientation of helimagnetic order with varying magnetic field. Depending on the history and the population of domains, this reorientation process might either correspond to a crossover, or involves a first-order or second-order phase transition at the critical field $H_{c1}$ \cite{Kataoka1981,Plumer_1981,Walker1989,Grigoriev2007,Bauer_2017}.

A quantitative theory of this reorientation process that is valid in the limit of small $\lambda_{\rm SOC}$ was recently presented by Bauer {\it et al.} and verified by detailed experiments on MnSi \cite{Bauer_2017}. With the help of dc and ac susceptibilities as well as neutron scattering experiments, the evolution of the helix orientation, specified by the unit vector $\hat Q(\vec H)$, was carefully tracked as a function of magnetic field for various field directions. The crystallographic $\langle 100\rangle$ direction plays a special role in that two subsequent $\mathds{Z}_2$ transitions could be observed confirming a theoretical prediction of Walker \cite{Walker1989}.

According to the theory of Ref.~\cite{Bauer_2017} the differential magnetic susceptibility $\partial_H M$ naturally decomposes into two parts. Whereas the first part derives from the helix with a fixed axis $\hat Q$, the second part is attributed to the field dependence of $\hat Q(\vec H)$. The reorientation $\hat Q(\vec H)$ is associated with large relaxation times $\tau$ because it requires the rotation of macroscopic helimagnetic domains. As a consequence, the ac susceptibility for frequencies $\omega \tau \gg 1$ is only sensitive to the first part, which was experimentally confirmed in Ref.~\cite{Bauer_2017} suggesting relaxation times exceeding seconds, $\tau \geq 1$ sec. 
Generally, the reorientation depends on the history of the sample due to different domain populations, for example, realized for finite- or zero-field cooling. In particular, hysteresis was found at the second-order phase transition at $H_{c1}$. The decrease of the field across $H_{c1}$ is accompanied with the formation of multiple domains. The coexistence of different domains within the sample might hamper the realization of the optimal trajectory $\hat Q(\vec H)$, especially, in the presence of long relaxation times $\tau$. As a result, distinct behavior can be observed upon increasing and decreasing the field across $H_{c1}$. 

In Ref.~\cite{Bauer_2017} only bulk probes were experimentally investigated so that the microscopic origin of the slow relaxation processes could not be identified. However, it was speculated that topological defects of the helimagnetic order, i.e., disclination and dislocations, might play a special role as they should naturally arise at the boundaries between different domains. A slow creep-like motion of dislocations was indeed identified by magnetic force microscopy (MFM) measurements on the surface of FeGe samples by Dussaux {\it et al.} \cite{Dussaux_2016} after the system had been quenched from the field-polarized state to $\vec H = 0$. The motion of dislocations during a MFM scan 
results in discontinuities of the helical pattern in the MFM image consisting of characteristic 180$^\circ$ phase shifts. Subsequently, it was also demonstrated both experimentally and theoretically that domain walls might comprise topological disclination and dislocation defects \cite{Schoenherr2018}. Nevertheless, a microscopic investigation of such relaxation events close to $H_{c1}$ has not been achieved so far. 

In the present work, we investigate the helix reorientation in the chiral magnet Cu\textsubscript{2}OSeO\textsubscript{3} using microscopic MFM measurements. This material is an insulator with a magnetoelectric coupling that allows to manipulate magnetic skyrmions and helices with electric fields, and it gives rise to various interesting magnetoelectric effects \cite{Seki2012,Mochizuki2015}. This material is also promising for magnonic applications due to its very low Gilbert damping parameter \cite{Garst_2017}. In constrast to MnSi, its helix is oriented along a $\langle 100 \rangle$ direction at zero field. The relatively large ratio $H_{c1}/H_{c2} \sim 0.36$ of Cu\textsubscript{2}OSeO\textsubscript{3} \cite{Halder2018} suggests that the spin-orbit coupling constant $\lambda_{\rm SOC}$ is larger than in MnSi. Indeed, additional magnetic phases stabilized by magnetocrystalline anisotropies -- the (metastable) canted conical state as well as the low temperature skyrmion lattice phase -- were found in Cu\textsubscript{2}OSeO\textsubscript{3} at low temperatures but for $\vec H$ only aligned along crystallographic $\langle 100\rangle$ directions \cite{Chacon:2018fe,Qianeaat7323,Halder2018}. Recently, real space observations addressing these states have been reported for a thin Cu\textsubscript{2}OSeO\textsubscript{3} lamella and field along a $\langle 100 \rangle$ direction \cite{Han_2020}.

In previous work \cite{Milde2016}, we have already investigated Cu\textsubscript{2}OSeO\textsubscript{3} with MFM at higher temperatures close to $T_c$ and identified all the magnetic phases, i.e., the helical and conical helimagnetic textures, the skyrmion lattice phase, and the field-polarized phase. Using Kelvin-probe force microscopy (KPFM) we determined the electric polarization and its field-dependence within these various phases. However, the reorientation process  was not addressed in Ref.~\cite{Milde2016} and it is at the focus of the present work.

Due to the restriction of our experimental setup, the magnetic field is always aligned perpendicular to the plane that is scanned by MFM, and for our sample probe this corresponds approximately to the crystallographic $[110]$ direction, $\vec H \parallel [110]$. We study the helix reorientation for this field direction and we determine the periodicity of the periodic surface pattern and its in-plane orientation. Assuming that the bulk helimagnetic order essentially extends towards the surface, we extract the orientation of the helix as a function of magnetic field. In addition, we determine the electric polarization and its behavior during the reorientation process. Our results are interpreted within the effective Landau theory of Ref.~\cite{Bauer_2017} that, strictly speaking, is only controlled for small $\lambda_{\rm SOC}$, and we find good agreement between theory and experiment. Moreover, we present microscopic evidence that the motion of dislocations along domain boundaries contributes to the magnetic relaxation close to the reorientation transition.

The structure of the paper is as follows. In section \ref{sec:ExpMethods} we present the experimental methods. In section \ref{sec:theory} we shortly review the theory of Ref.~\cite{Bauer_2017} and discuss its application to Cu\textsubscript{2}OSeO\textsubscript{3}. In particular, we point out the presence of a robust $\mathds Z_3$ transition for fields along $\langle 111 \rangle$. The theoretical prediction for the current experimental setup are presented and the electric polarization is evaluated as a function of the applied magnetic field. The experimental results are presented in section \ref{sec:expres}. From the MFM images we extract the orientation of the helimagnetic order and the presence of various domains as a function of magnetic field. The relaxation processes are shortly analysed, and the electric polarization is determined. Finally, we finish with a discussion of our results and a summary in section \ref{sec:discussion}.

\section{Experimental Methods}
\label{sec:ExpMethods}

We investigate the  same $(70\pm5)$~\textmu m thick plate sample with a polished crystallographic $(110)$ surface, as in our earlier work \cite{Milde2016}, where all details on the sample preparation can be found.
Choosing a lower temperature at $T = 10$~K compared to the former study ensured much slower dynamics and accessing a broader transition region enabled the detailed inspection of the reorientation of the helix axis as well as the observation of helical domains.   

For real-space imaging, we use magnetic force microscopy (MFM), that proved to be a valuable tool for studying complex spin textures such as magnetic bubble domains \cite{1995:Wadas:JApplPhys} or helices and skyrmions in helimagnets and magnetic thin films \cite{Milde2013, Kezsmarki2015, Zhang2016,BacaniHug_2016,Schoenherr2018,MasapoguPanagopoulos_2019}.
In the presence of an electric polarization, also electrostatic forces act on the MFM-tip.
Compensating these forces by means of Kelvin-probe force microscopy (KPFM) permits the detection of pristine MFM data while simultaneously revealing the contact potential difference $\Delta U_{\mathrm{cpd}}$ \cite{Weaver1991,Nonnenmacher1991,Zerweck2005}.
In turn, this reflects the shift of the electric potential induced by the magnetoelectric coupling.

MFM, KPFM and non-contact atomic force microscopy (nc-AFM) were performed in an Omicron cryogenic ultra-high vacuum STM/AFM instrument \cite{Omicron} using the RHK R9s electronics \cite{RHK} for scanning and data acquisition.
For all measurements, we used PPP-QMFMR probes from Nanosensors \cite{Nanosensors} driven at mechanical peak oscillation amplitudes of $A \approx 10$~nm.

MFM images were recorded in a two-step process. Firstly, the topography and the contact potential difference of the sample were recorded and the topographic 2D slope was canceled. Secondly, the MFM tip was retracted $20$~nm off the sample surface to record magnetic forces while scanning a plane above the sample surface. The KPFM-controller was switched off during this second step. After the first MFM image had been completed at $\pm 100$~mT, the magnetic field was changed automatically in constant steps of $4$~mT in between consecutive images. In order to ensure a correct compensation bias, we approached the tip to the sample before every field step and switched the KPFM-controller on. Note that the KPFM values change for every new magnetic field increment.
After the new field had been reached, we hold the KPFM-controller again constant and retracted the tip by the same lift height.
After the series of images had been completed a background image in the field-polarized state at $|\mu_0 H| = 250$~mT was taken.

\section{Theory}
\label{sec:theory}

The energy density for the magnetization $\vec M$ of cubic chiral magnets in the limit of small spin-orbit coupling $\lambda_{\rm SOC}$ is given by $\mathcal{E} = \mathcal{E}_0 + \mathcal{E}_{\rm dip} + \mathcal{E}_{\rm aniso}$ where 
\begin{align} \label{energy0}
\mathcal{E}_0 = \frac{J}{2} (\partial_i \vec M)^2 + \sigma D \vec M (\nabla \times \vec M) - \mu_0 \vec H \vec M
\end{align}
comprises the isotropic exchange interaction $J>0$, the Dzyaloshinskiii-Moriya interaction $D>0$ and the Zeeman energy. 
Depending on the chirality of the system, the sign $\sigma = \pm 1$. The competition between the first two terms results in spatially modulated magnetic order with a typical wavevector given by $Q = D/J$. The second term $\mathcal{E}_{\rm dip}$ contains the dipolar interaction, and the last term $\mathcal{E}_{\rm aniso}$ represents the magnetocrystalline anisotropies that are effectively small in the limit of small $\lambda_{\rm SOC}$. Under certain conditions, the magnetic helix minimizes the energy density $\mathcal{E}$ where its orientation is determined by both the Zeemann energy and the magnetocrystalline anisotropies $\mathcal{E}_{\rm aniso}$. In general, this leads to a helix reorientation as a function of applied magnetic field.

An effective theory for this helix reorientation was presented in Ref.~\cite{Bauer_2017} for MnSi. In section \ref{subsec:EffLandau} and \ref{subsec:transition} we review this theory for completeness and discuss its validity for Cu\textsubscript{2}OSeO\textsubscript{3}. In section \ref{subsec:expsetup} we focus on the theoretical predictions for the experimental setup. In section \ref{subsec:polarization} we present a theory for the electric polarization in Cu\textsubscript{2}OSeO\textsubscript{3} and its dependence on magnetic field.

\subsection{Effective Landau potential for the helix axis}
\label{subsec:EffLandau}

\begin{figure}
	\includegraphics[width=\columnwidth]{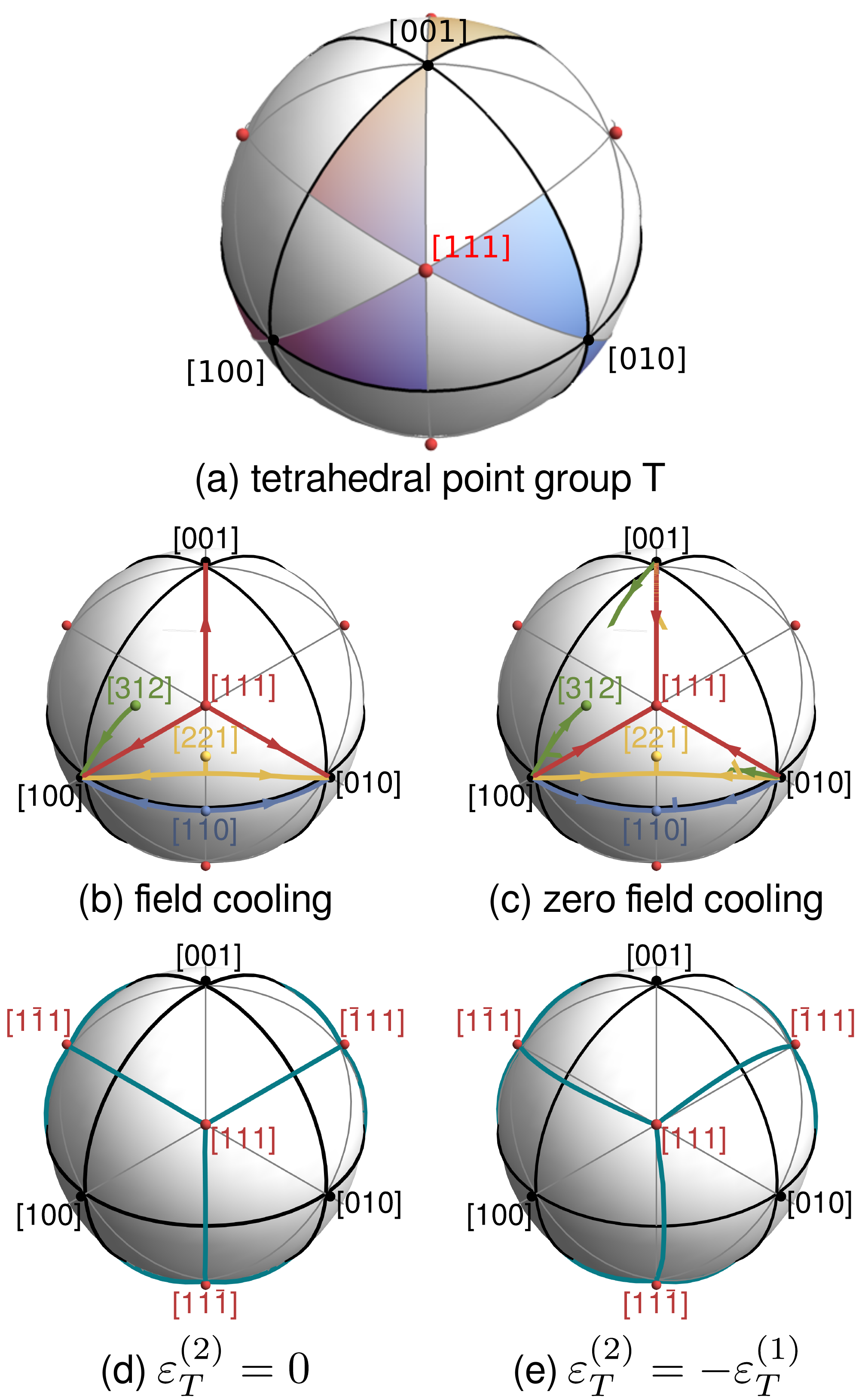}
	\caption{(a) B20 materials possess the symmetry of the tetrahedral point group indicated by the different colours on the sphere. (b) and (c): trajectories for the unit vector $\hat Q$ for different field directions for decreasing (field cooling) and increasing (zero field cooling) field magnitude, respectively, evaluated for $\varepsilon_T^{(2)} = 0$ in Eq.~\eqref{eq: V_T}. Starting from equally populated $\langle 100 \rangle$ domains at zero field in (c), trajectories can be continuous or discontinuous, and can possess a kink. (d) and (e): The kink represents a second-order transition that occur for magnetic field directions indicated by the blue solid lines. For a finite $\varepsilon_T^{(2)}$ these lines are warped towards $\langle 100\rangle$.} \label{fig:1}
\end{figure}

The helix wavevector $\vec Q$ is determined by the competition of Dzyaloshinskii-Moriya and exchange interaction and, as a consequence, its magnitude is proportional to spin-orbit coupling $|\vec Q| \sim \mathcal{O}(\lambda_{\rm SOC})$.
The orientation of the magnetic helix in a certain domain is represented in the following by the unit vector $\hat Q$. The competition between magnetocrystalline anisotropies and the Zeeman energy can be captured in the limit of small spin-orbit coupling $\lambda_{\rm SOC}$ by the Landau potential $\mathcal V(\hat Q) = \mathcal V_T(\hat Q)+ \mathcal V_H(\hat Q)$ \cite{Bauer_2017}. 

The first term represents the magnetocrystalline potential and its form is determined by the tetrahedral point group $T$ [see Fig.~\ref{fig:1}(a)] of B20 materials like MnSi or Cu\textsubscript{2}OSeO\textsubscript{3}. As indicated by the differently colored regions on the sphere, $T$ exhibits a three-fold $C_3$ rotation symmetry around the $\langle 111 \rangle$ directions, but only a two-fold $C_2$ rotation symmetry around $\langle 100 \rangle$ points. The potential $V_T(\hat Q)$ contains all terms consistent with $T$ and reads as
\begin{align}
\label{eq: V_T}
\mathcal V_T(\hat Q) &= \varepsilon_T^{(1)}(\hat Q_x^4 +\hat Q_y^4 + \hat Q_z^4)
\\ \notag
& 
+ \varepsilon_T^{(2)}\big(\hat Q_x^2 \hat Q_y^4+\hat Q_y^2 \hat Q_z^4+\hat Q_z^2 \hat Q_x^4\big)
\dots.
\end{align}
The lowest order term with constant $\varepsilon_T^{(1)} \sim \mathcal{O}(\lambda^4_{\rm SOC})$ is of fourth order in $\hat Q$, and it still possesses a four-fold rotation symmetry around the $\langle 100 \rangle$ axes that is not present in the tetrahedral point group $T$. This symmetry is broken explicitly only at the sixth order in $\hat Q$ by the term parameterized by $\varepsilon_T^{(2)} \sim \mathcal{O}(\lambda^6_{\rm SOC})$. Other terms of sixth and higher orders are represented by the dots. In the limit of small spin-orbit coupling, the first term determines the orientation of the helix at zero field. 
For $\varepsilon_T^{(1)} > 0$ the potential is minimized for $\hat Q \parallel \langle 111 \rangle$ as it is the case for MnSi whereas $\varepsilon_T^{(1)} < 0$ favours a helix orientation $\hat Q \parallel \langle 100 \rangle$ like in Cu\textsubscript{2}OSeO\textsubscript{3}. 

The second term in the Landau potential represents the Zeeman energy and up to second order in the applied magnetic field $\vec H$ it reads
\begin{equation}
\mathcal V_H(\hat Q) = -\frac{\mu_0}{2} H_i \chi_{ij} H_j,
\label{eq: V_H}
\end{equation}
where $\mu_0$ is the magnetic constant. The inverse of the magnetic susceptibility tensor evaluated at zero field is given by 
\begin{align}
\chi^{-1}_{ij} = \chi^{-1}_{ij,{\rm int}} + N_{ij},
\end{align}
with the demagnetization tensor $N_{ij}$ that is diagonal for an elliptical sample shape $N =$ diag$\{N_x,N_y,N_z\}$ with tr$\{N\} = 1$. The internal susceptibility tensor $\chi_{\rm int}$ is evaluated for a fixed orientation of the helimagnetic order and depends on $\hat Q$,
\begin{align}
\chi_{ij,{\rm int}} = \chi^{\rm int}_\parallel \hat Q_i \hat Q_j +\chi^{\rm int}_\perp (\delta_{ij}-\hat Q_i \hat Q_j).
\end{align}
An explicit calculation yields $\chi^{\rm int}_\parallel = 2\chi^{\rm int}_\perp$, i.e., only half of the spins respond to a transversal magnetic field compared to a field applied longitudinal to $\hat Q$. 

Minimization of the Landau potential $\mathcal V(\hat Q) = \mathcal V_T(\hat Q)+ \mathcal V_H(\hat Q)$ yields the helix orientation as a function of magnetic field $\hat Q = \hat Q(\vec H)$. The resulting trajectories were discussed in detail for $\varepsilon_T^{(1)} > 0$ in Ref.~\cite{Bauer_2017}. Here we focus on $\varepsilon_T^{(1)} < 0$. Next, we present a general discussion of the helix reorientation before turning to the configuration of the current experimental setup.

\subsection{Helix reorientation transitions}
\label{subsec:transition}

Depending on the direction of the applied magnetic field, the reorientation of the magnetic helix might involve either a crossover, a second-order phase transition or a first-order phase transition. For the purpose of a simplified discussion in this section we consider a sphere-like sample shape with demagnetization factors $N_i = 1/3$. First, we will focus on the potential without sixth-order terms and discuss corrections due to a finite $\varepsilon_T^{(2)}$ at the end of this section.
$\varepsilon_T^{(1)}$ in Cu\textsubscript{2}OSeO\textsubscript{3} is negative, i.e., the preferred directions in zero field are $\langle 100\rangle$ indicated by the black colored points on the unit sphere in Fig.~\ref{fig:1}. 

Figure \ref{fig:1}(b) presents trajectories of the helix axis $\hat Q(\vec H)$ for different field directions indicated by the coloured dots after field cooling. For high fields $\hat Q(\vec H) = \vec H/|\vec H|$. When decreasing the field, the crystalline anisotropies gain influence and the helix reorients towards $\langle 100 \rangle$. Depending on the field direction, three scenarios can be distinguished. The reorientation process is a crossover when the helix reorients smoothly towards the closest $\langle 100 \rangle$ direction like for the green trajectory in Fig.~\ref{fig:1}(b). It involves a second-order $\mathds Z_2$ transition when the trajectory bifurcates into two at a certain critical field $H_{c1}$, like for the blue and yellow trajectories. A special situation arises for a magnetic field along $\langle 111 \rangle$. Here, the trajectory can follow three  paths towards one of three distinct $\langle 100 \rangle$ directions realizing a second-order $\mathds Z_3$ transition. This transition is protected by the three-fold $C_3$ rotation symmetry of the point group $T$, i.e., it is robust even in the presence of a finite $\varepsilon_T^{(2)}$. In general, a $\mathds Z_3$ transition can be first-order as cubic terms are allowed in the effective Landau expansion. For the potential of Eq.~\eqref{eq: V_T} with $\varepsilon_T^{(2)} = 0$, however, this transition turns out to be of second-order with continuous trajectories $\hat Q(\vec H)$.

After zero field cooling, helimagnetic domains oriented along the three $\langle 100 \rangle$ directions are populated. Upon increasing the magnetic field, the helix axis $\hat Q$ moves towards the field direction [see Figure \ref{fig:1}(c)]. In addition to the reversed paths of panel (b) there exist also discontinuous paths starting from domains unfavoured by the field direction. This discontinuous reorientation correspond to a first-order transition.

The reorientation process thus involves a second-order transition and thus a well-defined critical field $H_{c1}$ only for specific directions of the magnetic field. For $\varepsilon_T^{(2)}=0$ these directions are located on the great circles on the sphere,that connect the $\langle 111 \rangle$ points [see Fig.~\ref{fig:1}(d)]. A finite $\varepsilon_T^{(2)}$ induces a warping of these lines [see panel (e)]. As the ratio $\varepsilon_T^{(2)}/\varepsilon_T^{(1)} \sim \lambda_{\rm SOC}^2$ is of second order in spin-orbit coupling this effect is expected to be small. 

In Cu\textsubscript{2}OSeO\textsubscript{3}, the sixth order term only quantitatively influences the reorientation transitions. This is different to MnSi where it is crucial to take the $\varepsilon_T^{(2)}$ term into account as discussed in Ref.~\cite{Bauer_2017}. There, $\varepsilon_T^{(1)}$ is positive which yields $\langle 111 \rangle$ as preferred directions in zero field. For a field along [100], four of those are equally close suggesting a $\mathds Z_4$ transition. However, a finite $\varepsilon_T^{(2)}$ splits this $\mathds Z_4$ transition into two subsequent $\mathds Z_2$ transitions.

\subsection{Helix orientation trajectory for the experimental setup}
\label{subsec:expsetup}

\begin{figure}
	\includegraphics[width=\columnwidth]{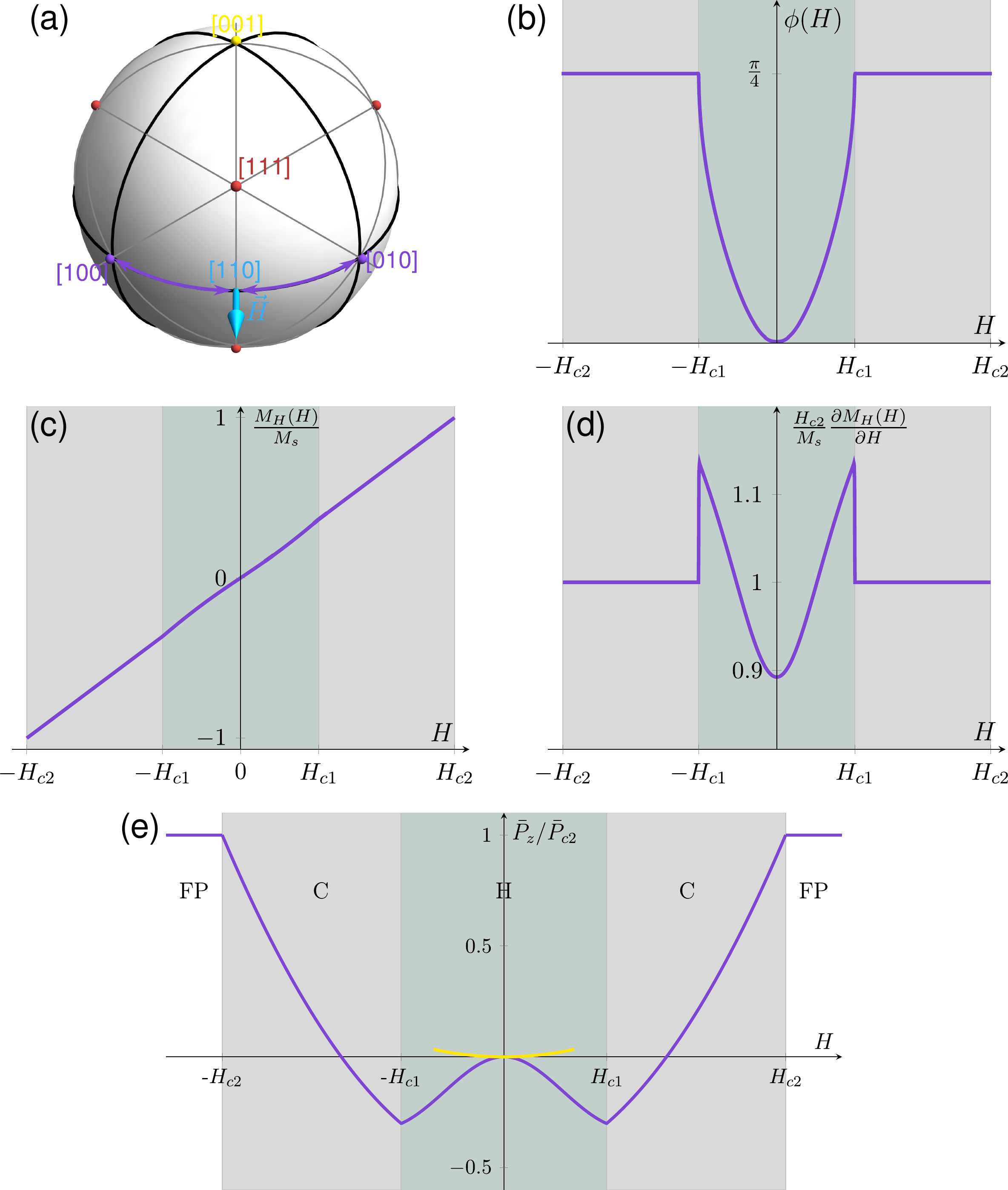}
	\caption{(a) Reorientation process for a magnetic field pointing along $[110]$. The helix axis $\hat Q = (\cos\phi,\sin\phi,0)$ continuously moves as a function of $H$ along the purple paths between $[110]$ and either $[100]$ or $[010]$. The (yellow) domain $[001]$ depopulates in a discontinuous fashion upon increasing $H$. (b) Azimuthal angle $\phi(H)$ starting from the $[100]$ domain at $H=0$; (c) uniform magnetization $M_H(H)$ and (d) susceptibility $\partial M_H/dH$
for the purple trajectories in (a). (e) Electrical polarization $\bar P_z (H)$ attributed to the domains located for $H=0$ at $[100]$ and $[010]$ (purple) and $[001]$ (yellow). The critical field $H_{c1}$ separates the conical state C (grey) where $\hat Q \parallel \vec H$ and the generalized helical state H (green) where $\hat Q \nparallel \vec H$.
	}
	\label{fig:2}
\end{figure} 

In the following, we neglect the sixth-order correction $\varepsilon_T^{(2)} = 0$ as its influence is weak and cannot be resolved within the experimental accuracy. The investigated sample [see section \ref{sec:ExpMethods}] is approximately a plate so that we use $N_x = N_y = 0$ and $N_z = 1$ for the demagnetization factors in the basis of principal axis. The normal axis of the plate-like sample approximately corresponds to a crystallographic $[110]$ direction. Within the crystallographic bases the demagnetization tensor is then given by 
\begin{equation}
\label{EqDemag}
N = \begin{pmatrix}
1/2 & 1/2 & 0\\
1/2 & 1/2 & 0\\
0 & 0 & 0\\
\end{pmatrix}.
\end{equation}
The magnetic field is always approximately aligned along the surface normal so that we restrict ourselves to the magnetic field direction $\hat H^T = (1,1,0)/\sqrt{2}$. For the longitudinal susceptibility of Cu\textsubscript{2}OSeO\textsubscript{3} we use the value $\chi^{\rm int}_\parallel = 1.76$ given in Ref.~\cite{2015:Schwarze:NatureMater}. The transition between the conical and field-polarized phase occurs at the critical field $\mu_0 H_{c2} \approx 192$ mT (see below), which we will use later on to normalize the field axis. 

For a field along $[110]$ the reorientation process is either first-order for the $[001]$ domain (yellow), or second-order for the $[100]$ and $[010]$ domains (purple) [see Fig.~\ref{fig:2}(a)]. The continuous trajectory of the helix axis can be parametrized as $\hat Q^T = (\cos \phi, \sin \phi, 0)$ with $\phi$ the azimuthal angle that depends on the magnetic field, $\phi = \phi(\vec H)$. Its dependence is shown in Fig.~\ref{fig:2}(b) where the kink defines the critical field, that for Eq.~\eqref{EqDemag} is given by
\begin{align}
\mu_0 H_{c1} = 2 (1+ \chi^{\rm int}_\parallel )\sqrt{ \frac{\mu_0 \varepsilon_T^{(1)}}{\chi^{\rm int}_\parallel}} .
\end{align}
As we will see later, experimentally we find $\mu_0 H_{c1} \approx 70$~mT corresponding to a value $\varepsilon_T^{(1)} = -0.0014$~$\mu$eV/\AA$^3$ for the magnetocrystalline potential. 

\begin{figure}[t!]
	\includegraphics[width=\columnwidth]{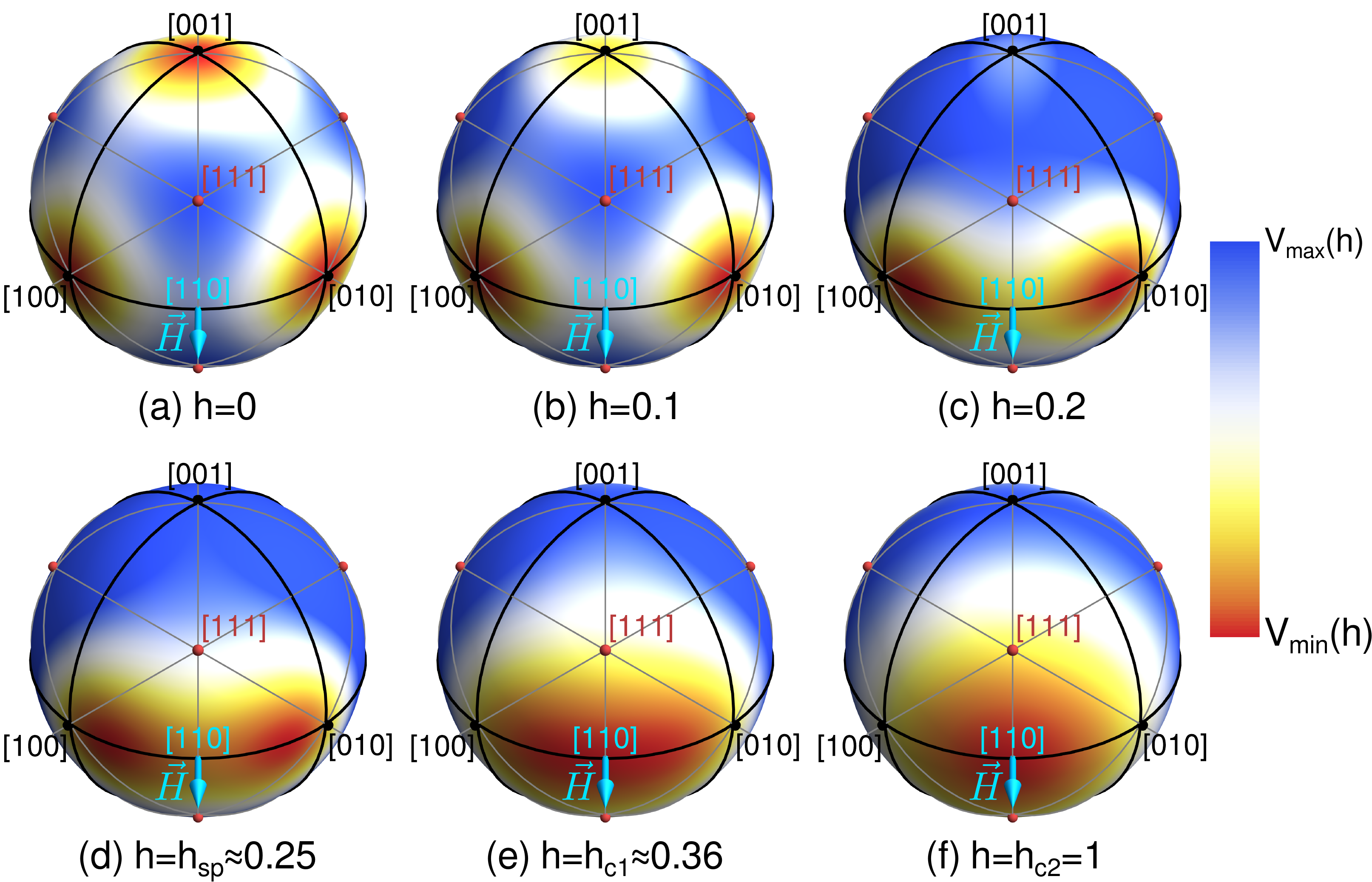}
	\caption{\label{fig:3}
	Effective Landau-potential for the helix axis, $\mathcal V(\hat Q)$, for the experimental parameters. The magnetic field is applied along $[110]$ and various values of the reduced field $h = H/H_{c2}$ are shown, see text. The color coding varies from panel to panel; red and blue color correspond to the minimal and maximal potential value at each specific field, respectively.
	}
\end{figure}

In Fig.~\ref{fig:3} we illustrate the Landau potential $\mathcal V(\hat Q)$ for the above parameters for various values of the magnetic field $H$. At zero field, all $\langle 100 \rangle$ directions are energetically degenerate, see panel (a). For a finite field along $[110]$, the direction $[001]$ remains a local minimum until it disappears at a certain spinodal field $H_{\rm sp}$ of $H_{\rm sp}/H_{c2} \approx 0.25$. At the same time, the other minima remain global minima and move towards the field direction. They merge at the critical field $H_{c1}/H_{c2} \approx 0.36$ and a single global minimum is obtained for $H \geq H_{c1}$.

\subsection{Electric polarization}
\label{subsec:polarization}

The magnetoelectric coupling in Cu$_2$OSeO$_3$ induces an electric polarization that is given in terms of the magnetization vector $\vec M$ by \cite{Seki2012}
\begin{align}
\vec P(\vec r\, ) = c_\mathrm{ME}\left\langle\left(\begin{matrix}
M_y(\vec r\, ) M_z(\vec r\, )\\
M_z(\vec r\, ) M_x(\vec r\, )\\
M_x(\vec r\, ) M_y(\vec r\, )
\end{matrix}\right)\right\rangle,
\label{eq: polarization <-> magnetization}
\end{align}
where $c_\mathrm{ME}$ denotes the magnetoeletric coupling constant. Generally, the expectation value in Eq.~\eqref{eq: polarization <-> magnetization} can be decomposed into
\begin{align}
\langle M_i(\vec r\, ) M_j(\vec r\, ) \rangle = \langle M_i(\vec r\,)\rangle \langle M_j(\vec r\,)\rangle + \mathcal{S}_{ij}(\vec r\,)
\end{align}
with $\mathcal{S}_{ij}(\vec r\,)$ the correlation function. In the mean-field approximation, $\mathcal{S}_{ij}$ is neglected and the polarization reduces to a product of expectation values $\langle \vec M(\vec r\,)\rangle$. 

Within the framework of the Landau theory of section \ref{subsec:EffLandau} this expectation value is given in terms of a Fourier series,
\begin{align} \label{Magnetization}
\langle \vec M(\vec r\,)\rangle = \vec M_H + \vec M_\mathrm{helix}(\vec r\,) + \delta \vec M(\vec r\,).
\end{align}
The second term is given by a harmonic helix 
\begin{align} \label{HarmHelix}
\vec M_\mathrm{helix}(\vec r\,) = M_\mathrm{helix} \Big[\hat e_1 \cos(\vec Q_\mathrm{min} \vec r) + \sigma \hat e_2 \sin(\vec Q_\mathrm{min} \vec r) \Big]
\end{align}
with the orthogonal unit vectors $\hat e_1 \times \hat e_2 = \hat Q_\mathrm{min} = \vec Q_\mathrm{min}/Q$. Depending on the chirality of the system, see Eq.~\eqref{energy0}, the helix can be right-handed or left-handed corresponding to $\sigma = +1$ or $-1$, respectively. Here, the orientation of the helix axis $\hat Q_\mathrm{min}(\vec H)$ minimizes the Landau potential $\mathcal V(\hat Q)$ at a given $\vec H$. The first term in Eq.~\eqref{Magnetization} represents the uniform part, $\vec M_H = M_H \hat H$, that can be obtained with the help of the Landau potential:
\begin{align}
M_H = -\frac{1}{\mu_0} \frac{\partial \mathcal V (\hat Q_\mathrm{min}(\vec H))}{\partial H}.
\end{align}
The magnitude $M_H$ as well as the total susceptibility $\partial_H M_H$ are  shown in Fig.~\ref{fig:2}(c) and (d) respectively. The susceptibility shows a pronounced mean-field jump at the critical field $H_{c1}$.

If variations of the amplitude are negligible, the length of the magnetization should be locally given by the saturation magnetization $\langle \vec M(\vec r\,)\rangle^2 = M_s^2$ which gives rise to anharmonicities represented by $\delta \vec M(\vec r\,)$ in Eq.~\eqref{Magnetization}. Minimizing the energy \eqref{energy0} in the presence of this constraint, we find in lowest order, i.e., neglecting $3 \vec Q$ Fourier components that $\delta \vec M$ also assumes the form of a helix,
\begin{align} \label{deltaM}
\delta \vec M(\vec r\,) \approx |\vec M_{H,\perp}|
\left[\hat e'_1 \cos(2 \vec Q_\mathrm{min} \vec r) + \sigma \hat e'_2 \sin(2 \vec Q_\mathrm{min} \vec r) \right].
\end{align}
The prefactor is determined by the magnetization projected onto the plane perpendicular to $\vec Q_{\rm min}$, i.e., $\vec M_{H,\perp} = \vec M_H - (\vec M_H \hat Q_{\rm min}) \hat Q_{\rm min}$. The unit vectors 
$\hat e'_1  \times \hat e'_2 = \hat Q_{\rm min}$ are given by
\begin{align}
\hat e'_1 &= \hat e_2 (\hat M_{H,\perp} \hat e_2) - \hat e_1 (\hat M_{H,\perp} \hat e_1),\\
\hat e'_2 &= - \hat e_2 (\hat M_{H,\perp} \hat e_1) - \hat e_1 (\hat M_{H,\perp} \hat e_2),
\end{align}
where $\hat M_{H,\perp} = \vec M_{H,\perp}/|\vec M_{H,\perp}|$. The component $\delta \vec M$ is proportional to the uniform magnetization $M_H$ and thus vanishes linearly with the applied magnetic field. Moreover, it vanishes for the conical state where $\vec M_H \parallel \hat Q_{\rm min}$ so that $\vec M_{H,\perp} = 0$. The anharmonicity is thus most pronounced at intermediate fields as observed in MnSi and FeGe \cite{Grigoriev_2006,Lebech_1989,Kousaka_2014_article}.
Within this approximation, the amplitude of the helix is given by 
\begin{align}
M_\mathrm{helix} = 
\sqrt{M_s^2 - \vec M_H^2 - \vec M^2_{H,\perp}}.
\end{align}

In the experiment, the polarization $\vec P(\vec r)$ cannot be spatially resolved on the scale of the helix wavelength. For this reason, we consider the polarization spatially averaged over a single period,
\begin{align}
\bar{\vec P} = \frac{1}{2\pi} \int\limits_0^{2\pi} dx\, \vec P(\vec r)\Big|^{\rm MF}_{x = \vec Q_{\rm min} \vec r}  
\end{align}
where the upper index MF indicates that we employ the mean-field approximation.
In our experimental setup, it turns out that only the z-component $\bar{P}_z$ is expected to remain finite. For a $(110)$ surface this $z$-component amounts to an in-plane polarization along $[001]$, $\bar{P}_{[001]} = \bar{P}_z$.

For the continuous trajectories, i.e., the purple paths of Fig.~\ref{fig:2}(a), we get
\begin{align} \label{Polarization_z}
\bar{P}_z = \frac{c_\mathrm{ME}}{2} \left[ M_H^2 - \frac{1}{2} \sin(2\phi) \left(M_s^2 - M_H^2\right) \right]\,,
\end{align}
with the azimuthal angle $\phi = \phi(H)$ of Fig.~\ref{fig:2}(b). 
Its magnetic field dependence corresponds to the purple line in Fig.~\ref{fig:2}(e). At the first critical field, $H_{c1}$, the polarization is minimal and shows a kink. For $H_{c2} > H >H_{c1}$, the angle $\phi = \pi/4$ and the uniform magnetization $M_H = M_s H / H_{c2}$ so that the polarization reduces to the known expression \cite{Seki2012,Aqeel2016,Milde2016} $\bar{P}_z = \frac{c_\mathrm{ME} M^2_s}{4} (3 (H/H_{c2})^2 - 1)$, and a sign change is expected for $H = H_{c2}/\sqrt{3}$. At the second critical field $H_{c2}$ another kink reflects the phase transition to the field-polarized phase. For $H \geq H_{c2}$, we have $M_H = M_s$ and $\bar{P}_z = \bar{P}_{c2} \equiv \frac{c_\mathrm{ME}}{2} M_s^2$. 

For the yellow $[001]$ domain in Fig.~\ref{fig:2}(a), the helix axis is given by $\hat Q^T_{\rm min} = (0,0,1)$ for fields $|H| \leq H_{\rm sp} \approx 0.25 H_{c2}$ up to its spinodal point where the first-order transition must take place at the latest. Its polarization within this field range is then given by 
\begin{align} \label{Polarization_z_2}
\bar{P}_z = \frac{c_\mathrm{ME}}{2} M_H^2,
\end{align}
that is shown as a yellow line in Fig.~\ref{fig:2}(e).

\begin{figure*}[tb!]
\includegraphics[width=\textwidth]{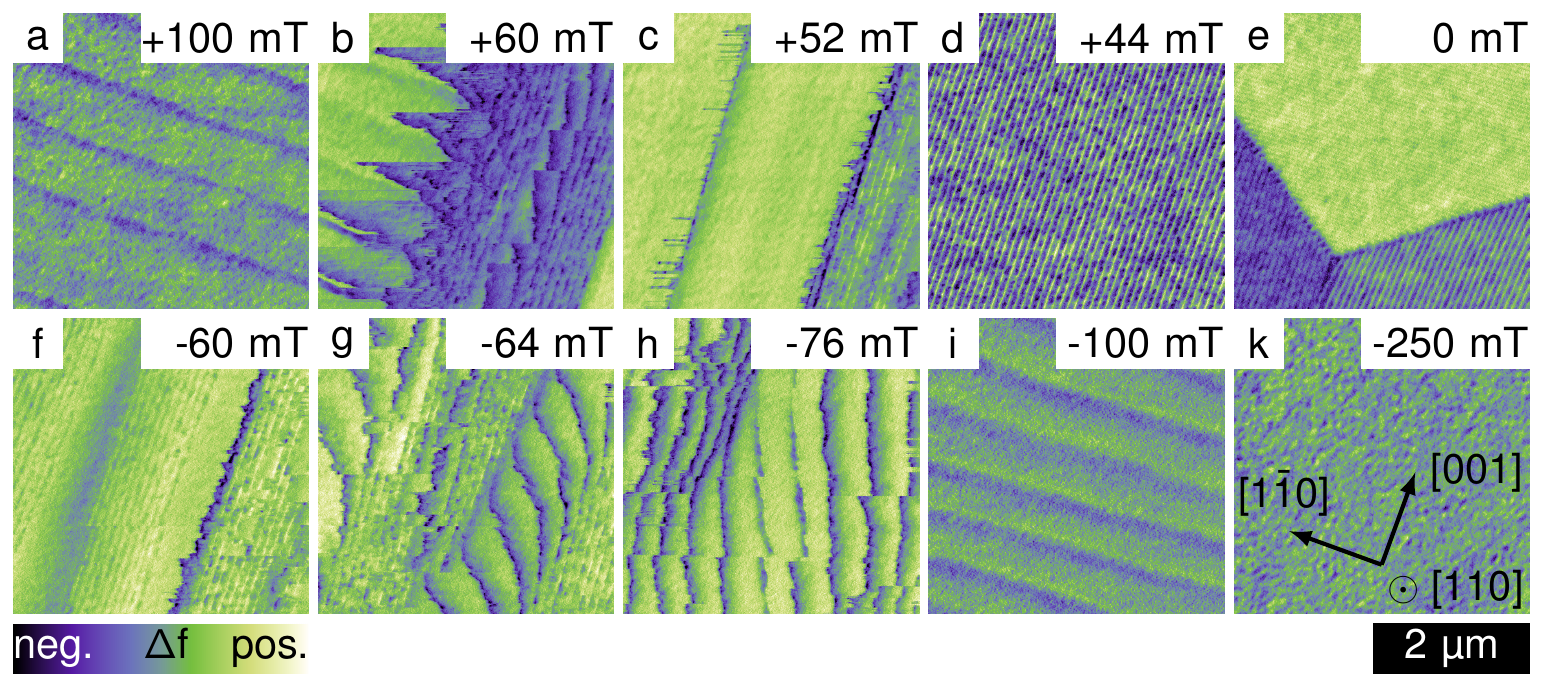}
\caption{ \label{fig:4} 
Typical scanning probe images of Cu\textsubscript{2}OSeO\textsubscript{3} at $T = 10$~K. 
The magnetic field was applied perpendicular to the plane of projection along the $[110]$-direction changing from positive to negative values. Periodic magnetic textures are observed whose wavelength and orientation depend on the strength of the magnetic field (see text). The span of the color scale is adapted individually, ranging from 1~Hz to 2.5~Hz.
}
\end{figure*} 

\section{Experimental results}
\label{sec:expres}

In this section, we present the experimental findings obtained via MFM and KPFM measurements that are both sensitive to signals attributed to the surface of the material. In the present setup an approximate $(110)$ surface is considered so that the surface normal $\hat n^T = (1,1,0)/\sqrt{2}$. Moreover, the applied field is approximately parallel to $\hat n$. It is important to note that helimagnetic order with orientation $\hat Q$ and an intrinsic wavelength $\lambda_h = 2\pi/Q \approx 60$ nm \cite{Adams2012} gives rise to periodic magnetic structures appearing at the sample surface characterized by a projected wavevector
\begin{equation}
\vec Q' = \vec Q - \hat n (\vec Q \hat n) = Q (\hat Q - \hat n \cos\Theta),
\end{equation}
where $\Theta   \in [0,\pi/2]$ is the angle between the helix axis $\hat Q$ and the surface normal $\hat n$. This results in a projected wavelength $\lambda^{\prime} = \frac{2\pi}{|\vec Q'|}$ given by \cite{Kezsmarki2015} 
\begin{equation} 
\lambda^{\prime} = \frac{\lambda_{h}}{\sin\Theta}. 
\label{eq1}
\end{equation} 
For a helix with an in-plane $\hat Q$ the angle $\Theta = \pi/2$ and $\lambda' = \lambda_h$. However, for a helix oriented along the surface normal $\Theta = 0$ the wavelength $\lambda'$ diverges, and the surface should appear homogeneous. 

Furthermore, for the later analysis we introduce the in-plane angle $\alpha = \sphericalangle(\hat e_{[001]}, \vec Q')$
defined as the angle enclosed by the projected wavevector $\vec Q'$ and the in-plane vector $\hat e^T_{[001]} = (0,0,1)$.

\subsection{Magnetic imaging of helimagnetic order}

We start the presentation of our experimental results with typical real-space images depicted in Fig.~\ref{fig:4}.
A slideshow of the complete dataset as well as of additional measurements can be retrieved from the supplementary materials.
As MFM essentially tracks the out-of-plane component of the local magnetization, the images represent the projection of the magnetization onto the surface normal $\vec M(\vec r)\hat n$.

The image series in Fig.~\ref{fig:4} is measured on the downwards branch of the hysteresis loop sweeping the external magnetic field from positive to negative values. After applying a saturating field of $\mu_0 H = 250$~mT,  the field was decreased and the series starts with $\mu_0 H = 100$~mT shown in Fig.~\ref{fig:4}(a) where a periodic pattern is visible. Decreasing the field further, the magnetization on the surface is reconstructed at about $\mu_0 H \approx 60$~mT and multiple helical domains form and increase in size preferentially showing a stripy pattern along the $[\bar{1}10]$ direction [see panel (c) and (d)]. Close to zero field, an additional helimagnetic domain oriented along the $[001]$-direction is observed giving rise to domain walls as shown in panel (e). The surface wavelength $\lambda'$ associated with the various domains depends on the applied magnetic field in a characteristic manner. When decreasing the field further to negative values, the domains start to split and magnetization reconstructs at about $\mu_0 H \approx - 70$~mT [see Fig.~\ref{fig:4}(g) and (h)]. At $\mu_0 H \approx -100 $~mT a periodic modulation oriented along $[001]$ is again visible in Fig.~\ref{fig:4}(i) similar to panel (a). Finally, at large field of $\mu_0 H= -250$~mT, the magnetization is fully polarized and the corresponding image in panel (k) is featureless.

We observed a manifold of co-existing domains as shown in Fig.~\ref{fig:4}(e) after field cycling. In contrast, after zero field cooling to $T = 10$ K only one single domain with an in-plane helix axis along $[001]$ could be observed, similarly to our previous measurements close to the critical temperature $T_c$ \cite{Milde2016}.

\subsection{Analysis of the MFM data}

\begin{figure}[t!]
	\includegraphics[width=\columnwidth]{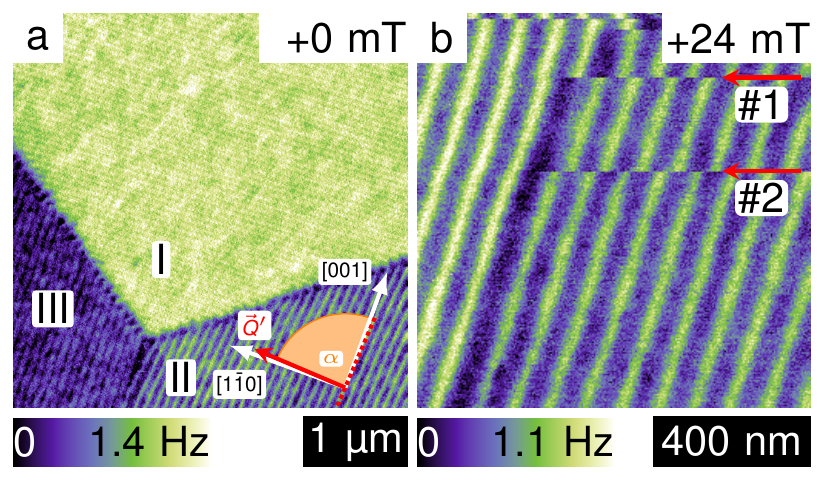}
	\caption{\label{fig:5} 
	(a) Close to field zero three types of domains, I, II, and III, can be observed. 
	The projected wavevector $\vec Q'$ and the $[001]$ direction enclose the in-plane angle $\alpha$. 
	(b) Relaxation events during the scanning process give rise to discontinuity lines (red arrows) within the periodic pattern with phase jumps of 180$^\circ$.}
\end{figure}

Assuming that the periodic patterns observed in the MFM images correspond to helimagnetic ordering projected onto the sample surface, we extract the projected wavevector $\vec Q'$, the projected wavelength $\lambda' = \frac{2\pi}{|\vec Q'|}$, and the corresponding angles $\Theta$ and $\alpha$ as defined at the beginning of section \ref{sec:expres}. Exemplified in Fig.~\ref{fig:5}(a), we can experimentally distinguish three types of domains depending on the in-plane angle $\alpha$ at zero field, namely type I with  $\alpha \sim 0$\textdegree, type II with $\alpha \lesssim 90$\textdegree, and type III with $\alpha \gtrsim 90$\textdegree. 

\begin{figure}[t!]
	\includegraphics[width=\columnwidth]{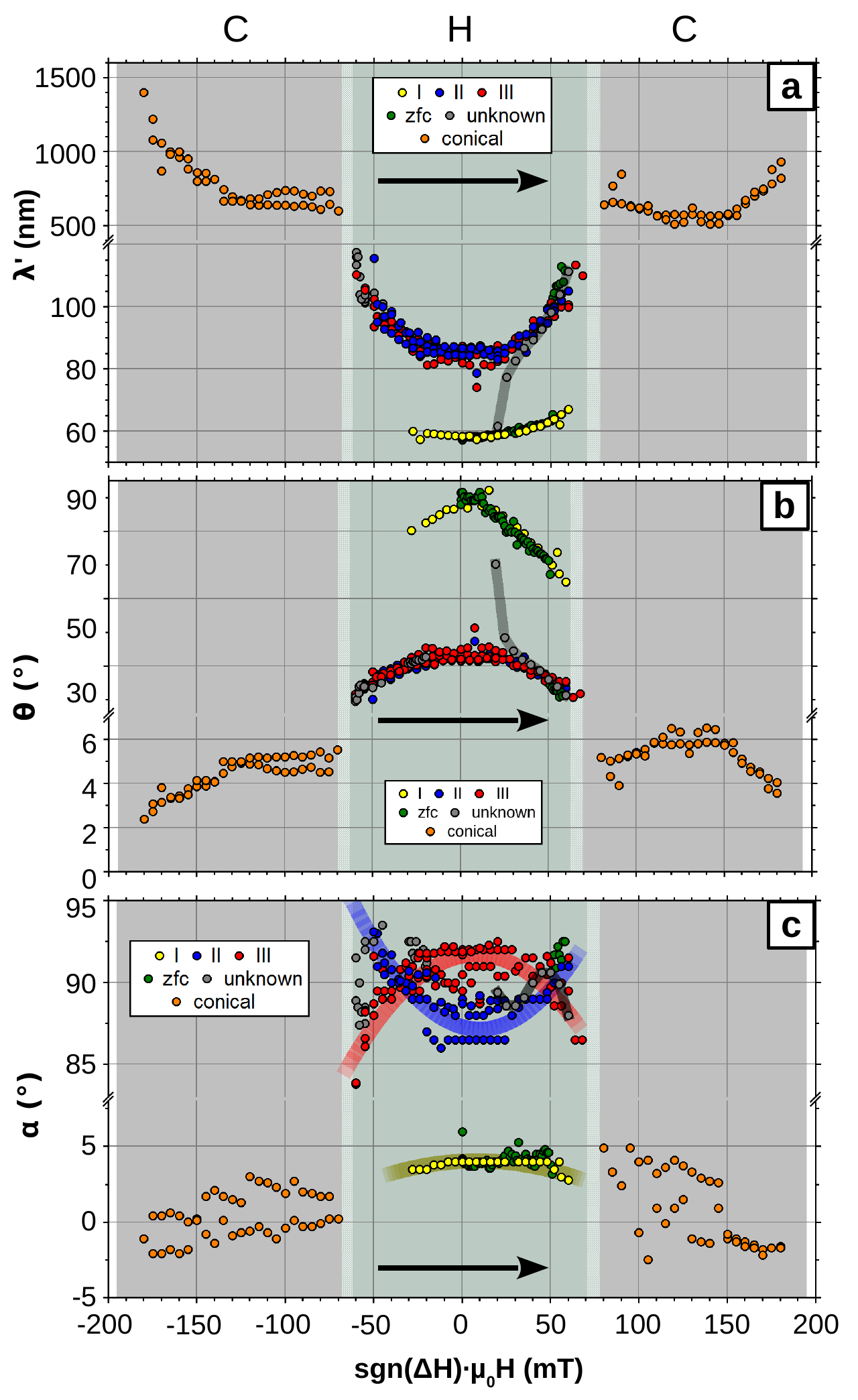}
	\caption{\label{fig:6} 
		Experimentally determined helix orientation characterized by (a) the projected wavelength $\lambda'$ of Eq.~\eqref{eq1}, (b) out-of-plane angle $\Theta = \sphericalangle(\hat n, \hat Q)$  and (c) measured in-plane angle $\alpha$.
		Yellow and green dots denote [001]-domains (type I) observed during field sweeps or after zero-field cooling, respectively.
		Blue and red dots denote type II domains and type III domains, respectively.
		Grey dots belong to domains, which where not in the image frame at zero field.}
\end{figure}

It was previously established by neutron scattering that the helices at zero field point along a crystallographic $\langle 100\rangle$ direction \cite{Adams2012}. Correspondingly, one expects indeed three different domains for a $(110)$ surface with in-plane angle $\alpha = 0$ for $\hat Q \parallel [001]$ and $\alpha = 90^\circ$ for $\hat Q$ along $[100]$ and $[010]$. The deviations from these values in Fig.~\ref{fig:6}(c) indicate an uncertainty of about $4^\circ$ due to a combination of systematic errors. First, the sample is slightly miscut so that the surface normal might be slightly tilted away from $[110]$ towards $[111]$, while second, the magnetic field might be slightly misaligned from the surface normal $\hat n$.
Third, the sample placement in the MFM can be slightly misaligned with a small in-plane rotation as well.
Finally, dynamic creep of the scanning piezo actuator slightly affects the scanner calibration. 

The evolution of the projected wavelength $\lambda'$ and the corresponding angle $\Theta$ of these three types of domains is shown as a function of magnetic field for every domain in Fig.~\ref{fig:6}(a) and (b), respectively. 
This includes also domains, which where not in the image frame at zero field and therefore are not classified as one of the three types in Fig.~\ref{fig:5}(a) (shown as grey dots).
For the helimagnetic domain oriented in-plane at zero field, $\lambda' = \lambda_h$ (yellow dots, green dots after zero-field cooling). The other domains (blue and red dots) are characterized for a $(110)$ surface by an angle $\Theta = \pi/4$ and a projected wavelength of $\lambda' = \sqrt{2} \lambda_h$. A drastic change of $\lambda'$ is observed around $70$ mT that we identify with the critical field $H_{c1}$ of the reorientation transition. For larger fields, the projected wavelength is of order $\lambda' \sim 10 \lambda_h$, corresponding to an angle $\Theta \sim 5^\circ$. We attribute this finite angle to the misalignment error mentioned above.

\subsection{Relaxation processes during helix reorientation}
\label{subsec:relax}

Whenever changing the magnetic field, the magnetic structure relaxes on relatively long time scales, especially close to $H_{c1}$ as discussed in Ref.~\cite{Bauer_2017}. An example of such a relaxation process is shown in Fig.~\ref{fig:5}(b). The MFM image is scanned from top to bottom. During this scan the magnetic structure might change due to relaxation events. They are reflected in discontinuity lines marked with (\#1) and (\#2) in Fig.~\ref{fig:5}(b), where the helix pattern is shifted by 180$^\circ$. 
Such 180$^\circ$ shifts were observed before in Ref.~\cite{Dussaux_2016} and attributed to the motion of dislocation defects in the helimagnetic background. Interestingly, the discontinuity lines do not continue through the full image frame but terminate. Probably, the termination points coincide with a helimagnetic domain wall  separating different $\langle 100\rangle$ domains \cite{Schoenherr2018}. This suggests that the discontinuity lines arise from motion of dislocations close to the domain wall. The creep motion of dislocations contributes to the complex and slow relaxation processes, giving rise to hysteretic effects even for the second-order phase transition at $H_{c1}$.

\subsection{Contact potential and polarization}

\begin{figure}
\includegraphics[width=\columnwidth]{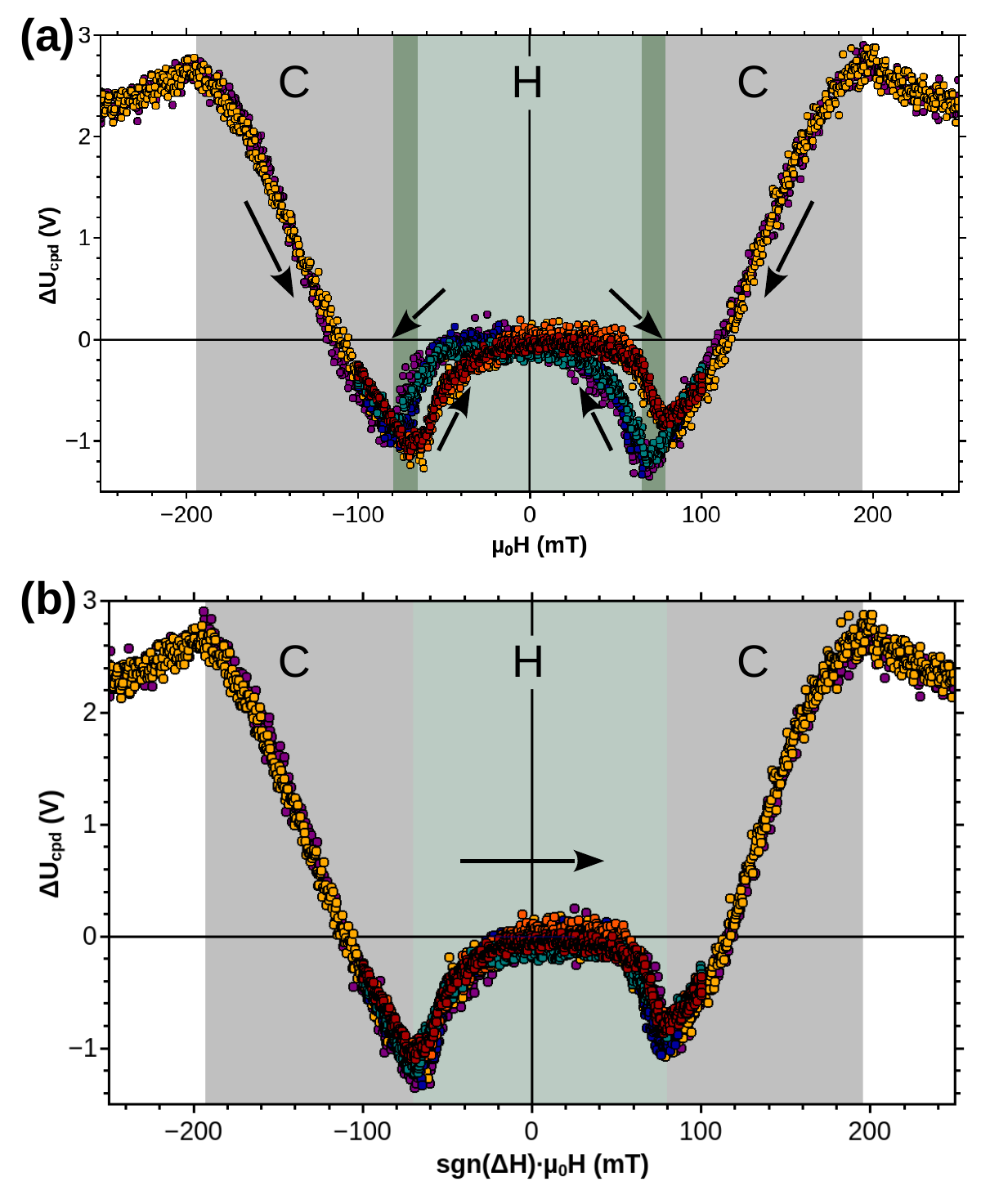}
\caption{\label{fig:7}  
(a) Contact potential difference $\Delta U_{\mathrm{cpd}}$ as a function of increasing (yellow, orange, and red points) and decreasing (green, blue and purple points) magnetic field at $T \approx 10$~K. 
(b) Same data is plotted so that all histories run from the left-hand to the right-hand side.
}
\end{figure}

Compensating electrostatic forces at the MFM tip with the help of KPFM allows to extract differences in the contact potential $\Delta U_{\rm cpd}$.
On a highly insulating sample as is Cu\textsubscript{2}OSeO\textsubscript{3}, this potential is measured on length scales of the mesoscopic MFM-tip so that it corresponds to an average over entire domains.
As explained in detail in Ref.~\cite{Milde2016}, for the current experimental setup this potential $\Delta U_{\rm cpd}$ for a single domain is proportional to the in-plane polarization $\bar{P}_z$ of Eqs.~\eqref{Polarization_z} or \eqref{Polarization_z_2}. The measured $\Delta U_{\rm cpd}$ as a function of magnetic field is displayed in Fig.~\ref{fig:7}(a) where the background colours indicate the various phases previously identified with MFM.

Similar to our previous measurement \cite{Milde2016} performed close to the critical temperature $T_c$, we find a plateau-like region close to the zero field, a minimum at the critical field $H_{c1}$, an increase of $\Delta U_{\rm cpd}$ within the conical phase $H_{c1} < H < H_{c2}$, and a kink at the second critical field $H_{c2}$. The main difference to the previous study in Ref.~\cite{Milde2016} is found at intermediate fields due to the absence of the skyrmion phase at 10 K in the present case. In addition, the hysteresis associated with the helix reorientation is more pronounced at lower temperatures due to slow relaxation processes already mentioned in section \ref{subsec:relax}. Performing a closed hysteresis loop, hysteretic effects are observed at both reorientation transitions $\pm H_{c1}$ [see Fig.~\ref{fig:7}(a)]. The hysteresis of both transitions is basically the same as illustrated in panel Fig.~\ref{fig:7}(b) where $\Delta U_{\rm cpd}$ has been replotted so that all histories run from left to right.  

\section{Discussion}
\label{sec:discussion}

\begin{figure}
	\centering
	\includegraphics[width=\columnwidth]{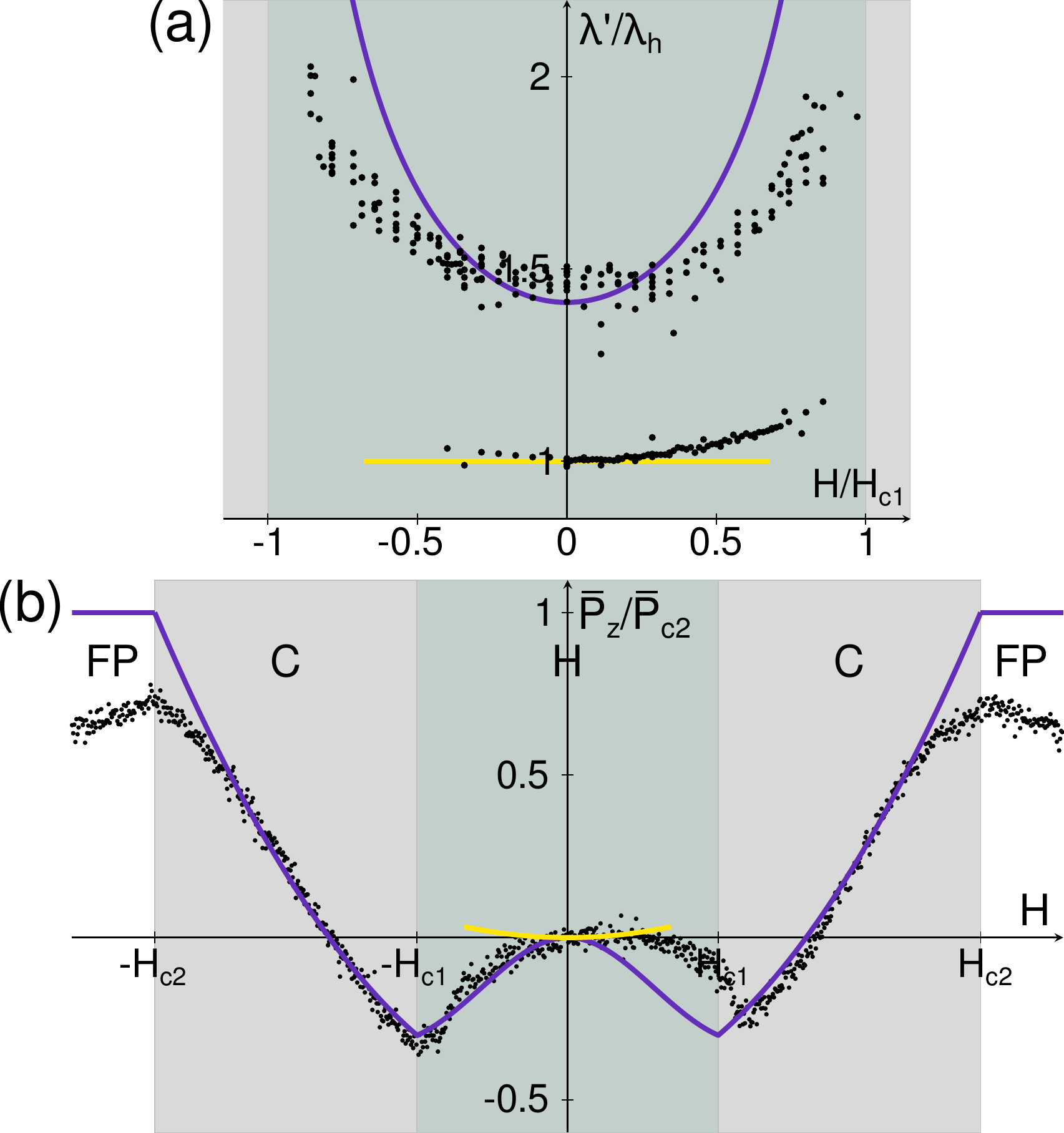}
	\caption{Comparison between experiment (dots) and theory (lines) for (a) the projected wavelength $\lambda'$ of Eq.~\eqref{eq1} and (b) the electric polarization as a function of magnetic field. The purple lines correspond to the contributions from domains starting in zero field at $[100]$ and $[010]$, and the yellow line corresponds to the $[001]$ domain [see Fig.~\ref{fig:2}(a)]. The experimental data were collected in a field sweep from negative to positive fields. } 
	\label{fig:8}
\end{figure}

In the following, our experimental results of section \ref{sec:expres} are interpreted in terms of the effective theory presented in section \ref{sec:theory}. As alluded to in section \ref{sec:expres}, the experimental system is plagued with systematic errors related to misalignments on the order of $5^\circ$. This will be reflected in the quantitative comparison to the theoretical predictions that presume the magnetic field being strictly aligned along $[110]$, the direction of the surface normal. The values for the critical fields were determined from characteristic kinks in the experimental data, $\mu_0 H_{c1} \approx 70 $ mT and $\mu_0 H_{c2} \approx 192 $ mT.  The latter value corresponds to an internal field of $\mu_0 H^{\rm int}_{c2} = \mu_0 (H_{c2} - N_z M_s) \approx 60$~mT for our plate-like sample $N_z = 0.92$ and the saturation magnetization $M_s \approx 110$~kA/m \cite{Halder2018, Bos2008}.

In Fig.~\ref{fig:8} we plot a comparison between theory and experiment for the projected wavelength $\lambda'$ of Eq.~\eqref{eq1} and Fig.~\ref{fig:6}(a) and the electric polarization of Fig.~\ref{fig:7}(b). The experimental data (dots) were collected in an up-sweep from negative to positive magnetic fields. The solid lines correspond to theoretical predictions where purple lines are attributed to domains located at $[100]$ and $[010]$ for $\vec H = 0$, and the yellow lines correspond to the in-plane $[001]$ domain. 
When the magnetic field is increased from negative values beyond $-H_{c1}$, the projected wavelength $\lambda'$ [see panel Fig.~\ref{fig:8}(a)], decreases in a characteristic fashion and achieves a minimum at $\vec H = 0$ before then increasing again in a similar manner on the other side. The theoretical curve reproduces this behavior qualitatively but overestimates the projected wavelength close to $H_{c1}$. Interestingly, the $[001]$ domain appears to be spontaneously populated upon approaching field zero, undergoing a first-order phase transition, although for the field direction $[110]$ this domain is always energetically unfavored at least within the bulk of the sample. As the helimagnetic axis of this  domain is located within the surface plane a projected wavelength $\lambda' = \lambda_h$ is expected that equals the wavelength within the bulk. This is indeed observed close to zero field, but $\lambda'$ slightly increases for increasing positive fields in contrast to the theoretical prediction (yellow line). 

In Fig.~\ref{fig:8}(b) the electric polarization is compared to theory. The measured polarization is spatially averaged over various domains so that it presents in general the combined signal from all three populated domains. Upon increasing the field from negative values, only two of the domains are populated so that the polarization closely follows the purple curve. However, when the $[001]$ domain gets spontaneously populated close to field zero, the corresponding polarization contributes possibly explaining the plateau-like feature of $\bar P_z$ for small positive fields. The behavior at the reorientation transition $H_{c1}$ is hysteretic and strongly depends on the history, which we attribute to a non-equilibrium effect similar to previous observations in MnSi \cite{Bauer_2017}.  Upon increasing the field towards $-H_{c1}$ the single helimagnetic domain splits into two, resulting in sharp signatures. However, increasing the field towards $+H_{c1}$, two or even three populated domains need to transform into a single domain which requires slow relaxation processes. The field-sweep employed in the experiment was probably too fast for the system to equilibrate giving rise to the apparent hysteretic behavior close to the continuous transition at $H_{c1}$. Close to the critical field $H_{c2}$, the measured polarization deviates from the theoretical prediction as it does not reach the same maximal value $\bar P_{c2}$ at $H_{c2}$. Furthermore, it decreases in the field-polarized phase instead of staying constant as theoretically expected. This disagreement probably is due to measuring artifacts at larger fields originating from a change in the distance between sample and tip.

In section \ref{subsec:relax} we provided experimental evidence for the involved relaxation processes. The 180$^\circ$  discontinuities observed in the MFM pattern hint at the motion of dislocations along domain boundaries [see Fig.~\ref{fig:5}(b)]. Similar 180$^\circ$ discontinuities have been previously observed in FeGe after a field quench \cite{Dussaux_2016}. 

The assumptions employed in the theoretical model of section \ref{sec:theory} need to be critically scrutinized. First of all, this model aimed at describing the reorientation of helimagnetic order within the bulk of the material. Additional contributions arising from the surface of the material can easily modify our assumption that the MFM images essentially reflect the projection of bulk helimagnetic order. It is known that so-called surface twists can lead to modification of helimagnetic order close to the surface \cite{Rybakov_2016}, which is neglected in the present work. In particular, such surface twists due to the magnetic boundary condition could lead to an anchoring of the helix axis within the surface plane. The boundary condition is automatically fulfilled for the pristine helix in case that the helix axis is aligned with the surface normal $\hat Q \parallel \hat n$. However, in the opposite limit $\hat Q \perp \hat n$ the boundary conditions will induce distortions to the helical texture, which might lower the surface energy of the magnetic structure and potentially favours surface domains with $\hat Q \perp \hat n$. This could explain the unexpected population of the $[001]$ domain with $\hat Q \perp \hat n$ observed in the present experiment for field-cooling with $\vec H \parallel [110]$. In addition, we exclusively observed $[001]$ helix configurations at higher temperatures \cite{Milde2016} as well as for zero-field cooling to $T = 10$ K. This suggests that the three $\langle 100 \rangle$ domains, which are degenerate within the bulk at $\vec H = 0$, are not evenly populated close to the surface so that the $\hat Q \perp \hat n$ surface domain is indeed energetically favoured. 

Furthermore, the theory presented is valid in the limit of small spin-orbit coupling $\lambda_{\rm SOC}$. However, this coupling is sufficiently strong in Cu\textsubscript{2}OSeO\textsubscript{3} so that additional phases are stabilized for magnetic fields along $\langle 100\rangle$ \cite{Chacon:2018fe,Qianeaat7323}. As discussed in Ref.~\cite{Halder2018}, the stabilization of a metastable tilted conical phase can be phenomenologically described in terms of a modified parameter $\varepsilon_T^{(1)}$ in Eq.~\eqref{eq: V_T} that is field dependent and changes sign as a function of $\vec H$. Whereas these effects are believed to be important mainly for fields along $\langle 100\rangle$, they might give rise to quantitative corrections for the present experimental setup with $\vec H \parallel [110]$.

We note that our theory of the electric polarization, $P_z$, for the reorientation process presented in section \ref{subsec:polarization} is also relevant for the spin-Hall magnetoresistance of Cu\textsubscript{2}OSeO\textsubscript{3} that was found to be proportional to $P_z$ \cite{Aqeel2016}.

In summary, we investigated the helix reorientation in Cu\textsubscript{2}OSeO\textsubscript{3} for a magnetic field aligned close to $[110]$ by means of magnetic force microscopy. This technique allows to probe the manifestations of the reorientation process at the sample surface with high spatial resolution. We observed the formation of domains close to the transition $H_{c1}$, and we identified relaxation events in real space that accompany the slow reorientation process.
Given the experimental uncertainties, the periodicity of the observed surface patterns are consistent with the projected wavelength of the bulk helimagnetic order, and its magnetic field dependence is in good agreement with theoretical predictions. Nevertheless, we also found evidence for a surface anchoring of the helix wavevector $\hat Q$ favouring domains with in-plane $\hat Q$. An interesting extension of the present work might be the detailed comparison between MFM and bulk measurements as well as the study of the reorientation process for fields along $\langle 111\rangle$ where a $\mathds{Z}_3$ transition is expected for Cu\textsubscript{2}OSeO\textsubscript{3}.

\section*{acknowledgement}
	
P.M., E.N., and L.M.E.\ gratefully acknowledge financial support by the German Science Foundation (DFG) through the Collaborative Research Center ``Correlated Magnetism: From Frustration to Topology'' (Project No. 247310070, Project C05), the SPP2137 (project no. EN~434/40-1), and project no. EN~434/38-1 and no. MI~2004/3-1. L.M.E. also gratefully acknowledges financial support through the Center of Excellence - Complexity and Topology in Quantum Matter (ct.qmat) - EXC 2147.
L.K.\ and M.G.\ gratefully acknowledge financial support by DFG SFB1143 (Correlated Magnetism: From Frustration To Topology, Project No. 247310070, Project A07), DFG Grant No. 1072/5-1 (Project No. 270344603), and DFG Grant No. 1072/6-1 (Project No. 324327023).
A.B.\ and C.P.\ acknowledge financial support by the Deutsche Forschungsgemeinschaft (DFG, German Research Foundation) under TRR80 (From Electronic Correlations to Functionality, Project No.\ 107745057, Project E1) and the excellence cluster MCQST under Germany's Excellence Strategy EXC-2111 (Project No.\ 390814868) as well as by the European Research Council (ERC) through Advanced Grants No.\ 291079 (TOPFIT) and No.\ 788031 (ExQuiSid) is gratefully acknowledged.

\bibliography{Bibliography}

\begin{thebibliography}{40}%
\makeatletter
\providecommand \@ifxundefined [1]{%
 \@ifx{#1\undefined}
}%
\providecommand \@ifnum [1]{%
 \ifnum #1\expandafter \@firstoftwo
 \else \expandafter \@secondoftwo
 \fi
}%
\providecommand \@ifx [1]{%
 \ifx #1\expandafter \@firstoftwo
 \else \expandafter \@secondoftwo
 \fi
}%
\providecommand \natexlab [1]{#1}%
\providecommand \enquote  [1]{``#1''}%
\providecommand \bibnamefont  [1]{#1}%
\providecommand \bibfnamefont [1]{#1}%
\providecommand \citenamefont [1]{#1}%
\providecommand \href@noop [0]{\@secondoftwo}%
\providecommand \href [0]{\begingroup \@sanitize@url \@href}%
\providecommand \@href[1]{\@@startlink{#1}\@@href}%
\providecommand \@@href[1]{\endgroup#1\@@endlink}%
\providecommand \@sanitize@url [0]{\catcode `\\12\catcode `\$12\catcode
  `\&12\catcode `\#12\catcode `\^12\catcode `\_12\catcode `\%12\relax}%
\providecommand \@@startlink[1]{}%
\providecommand \@@endlink[0]{}%
\providecommand \url  [0]{\begingroup\@sanitize@url \@url }%
\providecommand \@url [1]{\endgroup\@href {#1}{\urlprefix }}%
\providecommand \urlprefix  [0]{URL }%
\providecommand \Eprint [0]{\href }%
\providecommand \doibase [0]{http://dx.doi.org/}%
\providecommand \selectlanguage [0]{\@gobble}%
\providecommand \bibinfo  [0]{\@secondoftwo}%
\providecommand \bibfield  [0]{\@secondoftwo}%
\providecommand \translation [1]{[#1]}%
\providecommand \BibitemOpen [0]{}%
\providecommand \bibitemStop [0]{}%
\providecommand \bibitemNoStop [0]{.\EOS\space}%
\providecommand \EOS [0]{\spacefactor3000\relax}%
\providecommand \BibitemShut  [1]{\csname bibitem#1\endcsname}%
\let\auto@bib@innerbib\@empty
\bibitem [{\citenamefont {M\"uhlbauer}\ \emph {et~al.}(2009)\citenamefont
  {M\"uhlbauer}, \citenamefont {Binz}, \citenamefont {Jonietz}, \citenamefont
  {Pfleiderer}, \citenamefont {Rosch}, \citenamefont {Neubauer}, \citenamefont
  {Georgii},\ and\ \citenamefont {B\"oni}}]{Muehlbauer2009}%
  \BibitemOpen
  \bibfield  {author} {\bibinfo {author} {\bibfnamefont {S.}~\bibnamefont
  {M\"uhlbauer}}, \bibinfo {author} {\bibfnamefont {B.}~\bibnamefont {Binz}},
  \bibinfo {author} {\bibfnamefont {F.}~\bibnamefont {Jonietz}}, \bibinfo
  {author} {\bibfnamefont {C.}~\bibnamefont {Pfleiderer}}, \bibinfo {author}
  {\bibfnamefont {A.}~\bibnamefont {Rosch}}, \bibinfo {author} {\bibfnamefont
  {A.}~\bibnamefont {Neubauer}}, \bibinfo {author} {\bibfnamefont
  {R.}~\bibnamefont {Georgii}}, \ and\ \bibinfo {author} {\bibfnamefont
  {P.}~\bibnamefont {B\"oni}},\ }\bibfield  {title} {\enquote {\bibinfo {title}
  {Skyrmion lattice in a chiral magnet},}\ }\href@noop {} {\bibfield  {journal}
  {\bibinfo  {journal} {Science}\ }\textbf {\bibinfo {volume} {13}},\ \bibinfo
  {pages} {915} (\bibinfo {year} {2009})}\BibitemShut {NoStop}%
\bibitem [{\citenamefont {Grigoriev}\ \emph {et~al.}(2007)\citenamefont
  {Grigoriev}, \citenamefont {Dyadkin}, \citenamefont {Menzel}, \citenamefont
  {Schoenes}, \citenamefont {Chetverikov}, \citenamefont {Okorokov},
  \citenamefont {Eckerlebe},\ and\ \citenamefont {Maleyev}}]{Grigoriev2007}%
  \BibitemOpen
  \bibfield  {author} {\bibinfo {author} {\bibfnamefont {Sergey~V}\
  \bibnamefont {Grigoriev}}, \bibinfo {author} {\bibfnamefont {V~A}\
  \bibnamefont {Dyadkin}}, \bibinfo {author} {\bibfnamefont {D}~\bibnamefont
  {Menzel}}, \bibinfo {author} {\bibfnamefont {J}~\bibnamefont {Schoenes}},
  \bibinfo {author} {\bibfnamefont {Yu~O}\ \bibnamefont {Chetverikov}},
  \bibinfo {author} {\bibfnamefont {A~I}\ \bibnamefont {Okorokov}}, \bibinfo
  {author} {\bibfnamefont {H}~\bibnamefont {Eckerlebe}}, \ and\ \bibinfo
  {author} {\bibfnamefont {S~V}\ \bibnamefont {Maleyev}},\ }\bibfield  {title}
  {\enquote {\bibinfo {title} {{Magnetic structure of {F}e$_{1-x}${C}o$_x${S}i
  in a magnetic field studied via small-angle polarized neutron
  diffraction}},}\ }\href@noop {} {\bibfield  {journal} {\bibinfo  {journal}
  {Physical Review B}\ }\textbf {\bibinfo {volume} {76}},\ \bibinfo {pages}
  {224424} (\bibinfo {year} {2007})}\BibitemShut {NoStop}%
\bibitem [{\citenamefont {M\"unzer}\ \emph {et~al.}(2010)\citenamefont
  {M\"unzer}, \citenamefont {Neubauer}, \citenamefont {Adams}, \citenamefont
  {M\"uhlbauer}, \citenamefont {Franz}, \citenamefont {Jonietz}, \citenamefont
  {Georgii}, \citenamefont {B\"oni}, \citenamefont {Pedersen}, \citenamefont
  {Schmidt}, \citenamefont {Rosch},\ and\ \citenamefont
  {Pfleiderer}}]{Muenzer2010}%
  \BibitemOpen
  \bibfield  {author} {\bibinfo {author} {\bibfnamefont {W.}~\bibnamefont
  {M\"unzer}}, \bibinfo {author} {\bibfnamefont {A.}~\bibnamefont {Neubauer}},
  \bibinfo {author} {\bibfnamefont {T.}~\bibnamefont {Adams}}, \bibinfo
  {author} {\bibfnamefont {S.}~\bibnamefont {M\"uhlbauer}}, \bibinfo {author}
  {\bibfnamefont {C.}~\bibnamefont {Franz}}, \bibinfo {author} {\bibfnamefont
  {F.}~\bibnamefont {Jonietz}}, \bibinfo {author} {\bibfnamefont
  {R.}~\bibnamefont {Georgii}}, \bibinfo {author} {\bibfnamefont
  {P.}~\bibnamefont {B\"oni}}, \bibinfo {author} {\bibfnamefont
  {B.}~\bibnamefont {Pedersen}}, \bibinfo {author} {\bibfnamefont
  {M.}~\bibnamefont {Schmidt}}, \bibinfo {author} {\bibfnamefont
  {A.}~\bibnamefont {Rosch}}, \ and\ \bibinfo {author} {\bibfnamefont
  {C.}~\bibnamefont {Pfleiderer}},\ }\bibfield  {title} {\enquote {\bibinfo
  {title} {Skyrmion lattice in the doped semiconductor
  {F}e$_{1-x}${C}o$_x${S}i},}\ }\href@noop {} {\bibfield  {journal} {\bibinfo
  {journal} {Phys. Rev. B}\ }\textbf {\bibinfo {volume} {4}},\ \bibinfo {pages}
  {041203} (\bibinfo {year} {2010})}\BibitemShut {NoStop}%
\bibitem [{\citenamefont {Lebech}\ \emph {et~al.}(1989)\citenamefont {Lebech},
  \citenamefont {Bernhard},\ and\ \citenamefont {Freltoft}}]{Lebech_1989}%
  \BibitemOpen
  \bibfield  {author} {\bibinfo {author} {\bibfnamefont {B}~\bibnamefont
  {Lebech}}, \bibinfo {author} {\bibfnamefont {J}~\bibnamefont {Bernhard}}, \
  and\ \bibinfo {author} {\bibfnamefont {T}~\bibnamefont {Freltoft}},\
  }\bibfield  {title} {\enquote {\bibinfo {title} {Magnetic structures of cubic
  {F}e{G}e studied by small-angle neutron scattering},}\ }\href {\doibase
  10.1088/0953-8984/1/35/010} {\bibfield  {journal} {\bibinfo  {journal}
  {Journal of Physics: Condensed Matter}\ }\textbf {\bibinfo {volume} {1}},\
  \bibinfo {pages} {6105--6122} (\bibinfo {year} {1989})}\BibitemShut {NoStop}%
\bibitem [{\citenamefont {Yu}\ \emph {et~al.}(2010)\citenamefont {Yu},
  \citenamefont {Kanazawa}, \citenamefont {Onose}, \citenamefont {Kimoto},
  \citenamefont {Zhang}, \citenamefont {Ishiwata}, \citenamefont {Matsui},\
  and\ \citenamefont {Tokura}}]{Yu2010}%
  \BibitemOpen
  \bibfield  {author} {\bibinfo {author} {\bibfnamefont {X.Z.}\ \bibnamefont
  {Yu}}, \bibinfo {author} {\bibfnamefont {N.}~\bibnamefont {Kanazawa}},
  \bibinfo {author} {\bibfnamefont {Y.}~\bibnamefont {Onose}}, \bibinfo
  {author} {\bibfnamefont {K.}~\bibnamefont {Kimoto}}, \bibinfo {author}
  {\bibfnamefont {W.~Z.}\ \bibnamefont {Zhang}}, \bibinfo {author}
  {\bibfnamefont {S.}~\bibnamefont {Ishiwata}}, \bibinfo {author}
  {\bibfnamefont {Y.}~\bibnamefont {Matsui}}, \ and\ \bibinfo {author}
  {\bibfnamefont {Y.}~\bibnamefont {Tokura}},\ }\bibfield  {title} {\enquote
  {\bibinfo {title} {Near room-temperature formation of a skyrmion crystal in
  thin-films of the helimagnet {F}e{G}e},}\ }\href@noop {} {\bibfield
  {journal} {\bibinfo  {journal} {Nat. Mat.}\ }\textbf {\bibinfo {volume}
  {2}},\ \bibinfo {pages} {106} (\bibinfo {year} {2010})}\BibitemShut {NoStop}%
\bibitem [{\citenamefont {Seki}\ \emph
  {et~al.}(2012{\natexlab{a}})\citenamefont {Seki}, \citenamefont {Yu},
  \citenamefont {Ishiwata},\ and\ \citenamefont {Tokura}}]{Seki2012a}%
  \BibitemOpen
  \bibfield  {author} {\bibinfo {author} {\bibfnamefont {S.}~\bibnamefont
  {Seki}}, \bibinfo {author} {\bibfnamefont {X.Z.}\ \bibnamefont {Yu}},
  \bibinfo {author} {\bibfnamefont {S.}~\bibnamefont {Ishiwata}}, \ and\
  \bibinfo {author} {\bibfnamefont {Y.}~\bibnamefont {Tokura}},\ }\bibfield
  {title} {\enquote {\bibinfo {title} {Observation of skyrmions in a
  multiferroic material},}\ }\href@noop {} {\bibfield  {journal} {\bibinfo
  {journal} {Science}\ }\textbf {\bibinfo {volume} {336}},\ \bibinfo {pages}
  {198} (\bibinfo {year} {2012}{\natexlab{a}})}\BibitemShut {NoStop}%
\bibitem [{\citenamefont {Adams}\ \emph {et~al.}(2012)\citenamefont {Adams},
  \citenamefont {Chacon}, \citenamefont {Wagner}, \citenamefont {Bauer},
  \citenamefont {Brandl}, \citenamefont {Pedersen}, \citenamefont {Berger},
  \citenamefont {Lemmens},\ and\ \citenamefont {Pfleiderer}}]{Adams2012}%
  \BibitemOpen
  \bibfield  {author} {\bibinfo {author} {\bibfnamefont {T.}~\bibnamefont
  {Adams}}, \bibinfo {author} {\bibfnamefont {A.}~\bibnamefont {Chacon}},
  \bibinfo {author} {\bibfnamefont {M.}~\bibnamefont {Wagner}}, \bibinfo
  {author} {\bibfnamefont {A.}~\bibnamefont {Bauer}}, \bibinfo {author}
  {\bibfnamefont {G.}~\bibnamefont {Brandl}}, \bibinfo {author} {\bibfnamefont
  {B.}~\bibnamefont {Pedersen}}, \bibinfo {author} {\bibfnamefont
  {H.}~\bibnamefont {Berger}}, \bibinfo {author} {\bibfnamefont
  {P.}~\bibnamefont {Lemmens}}, \ and\ \bibinfo {author} {\bibfnamefont
  {C.}~\bibnamefont {Pfleiderer}},\ }\bibfield  {title} {\enquote {\bibinfo
  {title} {Long-wavelength helimagnetic order and skyrmion lattice phase in
  {C}u$_{2}${OS}e{O}$_{3}$},}\ }\href@noop {} {\bibfield  {journal} {\bibinfo
  {journal} {Phys. Rev. Lett.}\ }\textbf {\bibinfo {volume} {108}},\ \bibinfo
  {pages} {237204} (\bibinfo {year} {2012})}\BibitemShut {NoStop}%
\bibitem [{\citenamefont {Bauer}\ and\ \citenamefont
  {Pfleiderer}(2016)}]{BauerPfleiderer2016}%
  \BibitemOpen
  \bibfield  {author} {\bibinfo {author} {\bibfnamefont {Andreas}\ \bibnamefont
  {Bauer}}\ and\ \bibinfo {author} {\bibfnamefont {Christian}\ \bibnamefont
  {Pfleiderer}},\ }\enquote {\bibinfo {title} {Generic aspects of skyrmion
  lattices in chiral magnets},}\ in\ \href {\doibase
  10.1007/978-3-319-25301-5_2} {\emph {\bibinfo {booktitle} {Topological
  Structures in Ferroic Materials: Domain Walls, Vortices and Skyrmions}}},\
  \bibinfo {editor} {edited by\ \bibinfo {editor} {\bibfnamefont {Jan}\
  \bibnamefont {Seidel}}}\ (\bibinfo  {publisher} {Springer International
  Publishing},\ \bibinfo {year} {2016})\ pp.\ \bibinfo {pages}
  {1--28}\BibitemShut {NoStop}%
\bibitem [{\citenamefont {Kataoka}\ and\ \citenamefont
  {Nakanishi}(1981)}]{Kataoka1981}%
  \BibitemOpen
  \bibfield  {author} {\bibinfo {author} {\bibfnamefont {Mitsuo}\ \bibnamefont
  {Kataoka}}\ and\ \bibinfo {author} {\bibfnamefont {Osamu}\ \bibnamefont
  {Nakanishi}},\ }\bibfield  {title} {\enquote {\bibinfo {title} {Helical spin
  density wave due to antisymmetric exchange interaction},}\ }\href {\doibase
  10.1143/JPSJ.50.3888} {\bibfield  {journal} {\bibinfo  {journal} {Journal of
  the Physical Society of Japan}\ }\textbf {\bibinfo {volume} {50}},\ \bibinfo
  {pages} {3888--3896} (\bibinfo {year} {1981})}\BibitemShut {NoStop}%
\bibitem [{\citenamefont {Plumer}\ and\ \citenamefont
  {Walker}(1981)}]{Plumer_1981}%
  \BibitemOpen
  \bibfield  {author} {\bibinfo {author} {\bibfnamefont {M~L}\ \bibnamefont
  {Plumer}}\ and\ \bibinfo {author} {\bibfnamefont {M~B}\ \bibnamefont
  {Walker}},\ }\bibfield  {title} {\enquote {\bibinfo {title} {Wavevector and
  spin reorientation in {{M}n{S}i}},}\ }\href {\doibase
  10.1088/0022-3719/14/31/016} {\bibfield  {journal} {\bibinfo  {journal}
  {Journal of Physics C: Solid State Physics}\ }\textbf {\bibinfo {volume}
  {14}},\ \bibinfo {pages} {4689--4699} (\bibinfo {year} {1981})}\BibitemShut
  {NoStop}%
\bibitem [{\citenamefont {Walker}(1989)}]{Walker1989}%
  \BibitemOpen
  \bibfield  {author} {\bibinfo {author} {\bibfnamefont {M.~B.}\ \bibnamefont
  {Walker}},\ }\bibfield  {title} {\enquote {\bibinfo {title} {Phason
  instabilities and successive wave-vector reorientation phase transitions in
  {M}n{S}i},}\ }\href {\doibase 10.1103/PhysRevB.40.9315} {\bibfield  {journal}
  {\bibinfo  {journal} {Phys. Rev. B}\ }\textbf {\bibinfo {volume} {40}},\
  \bibinfo {pages} {9315--9317} (\bibinfo {year} {1989})}\BibitemShut {NoStop}%
\bibitem [{\citenamefont {Bauer}\ \emph {et~al.}(2017)\citenamefont {Bauer},
  \citenamefont {Chacon}, \citenamefont {Wagner}, \citenamefont {Halder},
  \citenamefont {Georgii}, \citenamefont {Rosch}, \citenamefont {Pfleiderer},\
  and\ \citenamefont {Garst}}]{Bauer_2017}%
  \BibitemOpen
  \bibfield  {author} {\bibinfo {author} {\bibfnamefont {A.}~\bibnamefont
  {Bauer}}, \bibinfo {author} {\bibfnamefont {A.}~\bibnamefont {Chacon}},
  \bibinfo {author} {\bibfnamefont {M.}~\bibnamefont {Wagner}}, \bibinfo
  {author} {\bibfnamefont {M.}~\bibnamefont {Halder}}, \bibinfo {author}
  {\bibfnamefont {R.}~\bibnamefont {Georgii}}, \bibinfo {author} {\bibfnamefont
  {A.}~\bibnamefont {Rosch}}, \bibinfo {author} {\bibfnamefont
  {C.}~\bibnamefont {Pfleiderer}}, \ and\ \bibinfo {author} {\bibfnamefont
  {M.}~\bibnamefont {Garst}},\ }\bibfield  {title} {\enquote {\bibinfo {title}
  {Symmetry breaking, slow relaxation dynamics, and topological defects at the
  field-induced helix reorientation in {M}n{S}i},}\ }\href@noop {} {\bibfield
  {journal} {\bibinfo  {journal} {Phys. Rev. B}\ }\textbf {\bibinfo {volume}
  {95}},\ \bibinfo {pages} {024429} (\bibinfo {year} {2017})}\BibitemShut
  {NoStop}%
\bibitem [{\citenamefont {Dussaux}\ \emph {et~al.}(2016)\citenamefont
  {Dussaux}, \citenamefont {Schoenherr}, \citenamefont {Koumpouras},
  \citenamefont {Chico}, \citenamefont {Chang}, \citenamefont {Lorenzelli},
  \citenamefont {Kanazawa}, \citenamefont {Tokura}, \citenamefont {Garst},
  \citenamefont {Bergman}, \citenamefont {Degen},\ and\ \citenamefont
  {Meier}}]{Dussaux_2016}%
  \BibitemOpen
  \bibfield  {author} {\bibinfo {author} {\bibfnamefont {A.}~\bibnamefont
  {Dussaux}}, \bibinfo {author} {\bibfnamefont {P.}~\bibnamefont {Schoenherr}},
  \bibinfo {author} {\bibfnamefont {K.}~\bibnamefont {Koumpouras}}, \bibinfo
  {author} {\bibfnamefont {J.}~\bibnamefont {Chico}}, \bibinfo {author}
  {\bibfnamefont {K.}~\bibnamefont {Chang}}, \bibinfo {author} {\bibfnamefont
  {L.}~\bibnamefont {Lorenzelli}}, \bibinfo {author} {\bibfnamefont
  {N.}~\bibnamefont {Kanazawa}}, \bibinfo {author} {\bibfnamefont
  {Y.}~\bibnamefont {Tokura}}, \bibinfo {author} {\bibfnamefont
  {M.}~\bibnamefont {Garst}}, \bibinfo {author} {\bibfnamefont
  {A.}~\bibnamefont {Bergman}}, \bibinfo {author} {\bibfnamefont {C.~L.}\
  \bibnamefont {Degen}}, \ and\ \bibinfo {author} {\bibfnamefont
  {D.}~\bibnamefont {Meier}},\ }\bibfield  {title} {\enquote {\bibinfo {title}
  {Local dynamics of topological magnetic defects in the itinerant helimagnet
  {F}e{G}e},}\ }\href {\doibase https://doi.org/10.1038/ncomms12430
  10.1038/ncomms12430} {\bibfield  {journal} {\bibinfo  {journal} {Nature
  Communications}\ }\textbf {\bibinfo {volume} {7}},\ \bibinfo {pages} {12430}
  (\bibinfo {year} {2016})}\BibitemShut {NoStop}%
\bibitem [{\citenamefont {Schoenherr}\ \emph {et~al.}(2018)\citenamefont
  {Schoenherr}, \citenamefont {M\"uller}, \citenamefont {K\"ohler},
  \citenamefont {Rosch}, \citenamefont {Kanazawa}, \citenamefont {Tokura},
  \citenamefont {Garst},\ and\ \citenamefont {Meier}}]{Schoenherr2018}%
  \BibitemOpen
  \bibfield  {author} {\bibinfo {author} {\bibfnamefont {P.}~\bibnamefont
  {Schoenherr}}, \bibinfo {author} {\bibfnamefont {J.}~\bibnamefont
  {M\"uller}}, \bibinfo {author} {\bibfnamefont {L.}~\bibnamefont {K\"ohler}},
  \bibinfo {author} {\bibfnamefont {A.}~\bibnamefont {Rosch}}, \bibinfo
  {author} {\bibfnamefont {N.}~\bibnamefont {Kanazawa}}, \bibinfo {author}
  {\bibfnamefont {Y.}~\bibnamefont {Tokura}}, \bibinfo {author} {\bibfnamefont
  {M.}~\bibnamefont {Garst}}, \ and\ \bibinfo {author} {\bibfnamefont
  {D.}~\bibnamefont {Meier}},\ }\bibfield  {title} {\enquote {\bibinfo {title}
  {Topological domain walls in helimagnets},}\ }\href@noop {} {\bibfield
  {journal} {\bibinfo  {journal} {Nat. Phys.}\ }\textbf {\bibinfo {volume}
  {14}},\ \bibinfo {pages} {465} (\bibinfo {year} {2018})}\BibitemShut
  {NoStop}%
\bibitem [{\citenamefont {Seki}\ \emph
  {et~al.}(2012{\natexlab{b}})\citenamefont {Seki}, \citenamefont {Ishiwata},\
  and\ \citenamefont {Tokura}}]{Seki2012}%
  \BibitemOpen
  \bibfield  {author} {\bibinfo {author} {\bibfnamefont {S.}~\bibnamefont
  {Seki}}, \bibinfo {author} {\bibfnamefont {S.}~\bibnamefont {Ishiwata}}, \
  and\ \bibinfo {author} {\bibfnamefont {Y.}~\bibnamefont {Tokura}},\
  }\bibfield  {title} {\enquote {\bibinfo {title} {Magnetoelectric nature of
  skyrmions in a chiral magnetic insulator {C}u$_{2}${OS}e{O}$_{3}$},}\
  }\href@noop {} {\bibfield  {journal} {\bibinfo  {journal} {Phys. Rev. B}\
  }\textbf {\bibinfo {volume} {86}},\ \bibinfo {pages} {060403(R)} (\bibinfo
  {year} {2012}{\natexlab{b}})}\BibitemShut {NoStop}%
\bibitem [{\citenamefont {Mochizuki}\ and\ \citenamefont
  {Seki}(2015)}]{Mochizuki2015}%
  \BibitemOpen
  \bibfield  {author} {\bibinfo {author} {\bibfnamefont {M.}~\bibnamefont
  {Mochizuki}}\ and\ \bibinfo {author} {\bibfnamefont {S.}~\bibnamefont
  {Seki}},\ }\bibfield  {title} {\enquote {\bibinfo {title} {Dynamical
  magnetoelectric phenomena of multiferroic skyrmions},}\ }\href@noop {}
  {\bibfield  {journal} {\bibinfo  {journal} {J. Phys.: Condens. Matter}\
  }\textbf {\bibinfo {volume} {27}},\ \bibinfo {pages} {503001} (\bibinfo
  {year} {2015})}\BibitemShut {NoStop}%
\bibitem [{\citenamefont {Garst}\ \emph {et~al.}(2017)\citenamefont {Garst},
  \citenamefont {Waizner},\ and\ \citenamefont {Grundler}}]{Garst_2017}%
  \BibitemOpen
  \bibfield  {author} {\bibinfo {author} {\bibfnamefont {Markus}\ \bibnamefont
  {Garst}}, \bibinfo {author} {\bibfnamefont {Johannes}\ \bibnamefont
  {Waizner}}, \ and\ \bibinfo {author} {\bibfnamefont {Dirk}\ \bibnamefont
  {Grundler}},\ }\bibfield  {title} {\enquote {\bibinfo {title} {Collective
  spin excitations of helices and magnetic skyrmions: review and perspectives
  of magnonics in non-centrosymmetric magnets},}\ }\href@noop {} {\bibfield
  {journal} {\bibinfo  {journal} {J. of Phys. D: Appl. Phys.}\ }\textbf
  {\bibinfo {volume} {50}},\ \bibinfo {pages} {293002} (\bibinfo {year}
  {2017})}\BibitemShut {NoStop}%
\bibitem [{\citenamefont {Halder}\ \emph {et~al.}(2018)\citenamefont {Halder},
  \citenamefont {Chacon}, \citenamefont {Bauer}, \citenamefont {Simeth},
  \citenamefont {M\"uhlbauer}, \citenamefont {Berger}, \citenamefont {Heinen},
  \citenamefont {Garst}, \citenamefont {Rosch},\ and\ \citenamefont
  {Pfleiderer}}]{Halder2018}%
  \BibitemOpen
  \bibfield  {author} {\bibinfo {author} {\bibfnamefont {M.}~\bibnamefont
  {Halder}}, \bibinfo {author} {\bibfnamefont {A.}~\bibnamefont {Chacon}},
  \bibinfo {author} {\bibfnamefont {A.}~\bibnamefont {Bauer}}, \bibinfo
  {author} {\bibfnamefont {W.}~\bibnamefont {Simeth}}, \bibinfo {author}
  {\bibfnamefont {S.}~\bibnamefont {M\"uhlbauer}}, \bibinfo {author}
  {\bibfnamefont {H.}~\bibnamefont {Berger}}, \bibinfo {author} {\bibfnamefont
  {L.}~\bibnamefont {Heinen}}, \bibinfo {author} {\bibfnamefont
  {M.}~\bibnamefont {Garst}}, \bibinfo {author} {\bibfnamefont
  {A.}~\bibnamefont {Rosch}}, \ and\ \bibinfo {author} {\bibfnamefont
  {C.}~\bibnamefont {Pfleiderer}},\ }\bibfield  {title} {\enquote {\bibinfo
  {title} {Thermodynamic evidence of a second skyrmion lattice phase and tilted
  conical phase in {C}u$_{2}${OS}e{O}$_{3}$},}\ }\href {\doibase
  10.1103/PhysRevB.98.144429} {\bibfield  {journal} {\bibinfo  {journal} {Phys.
  Rev. B}\ }\textbf {\bibinfo {volume} {98}},\ \bibinfo {pages} {144429}
  (\bibinfo {year} {2018})}\BibitemShut {NoStop}%
\bibitem [{\citenamefont {Chacon}\ \emph {et~al.}(2018)\citenamefont {Chacon},
  \citenamefont {Heinen}, \citenamefont {Halder}, \citenamefont {Bauer},
  \citenamefont {Simeth}, \citenamefont {M{\"u}hlbauer}, \citenamefont
  {Berger}, \citenamefont {Garst}, \citenamefont {Rosch},\ and\ \citenamefont
  {Pfleiderer}}]{Chacon:2018fe}%
  \BibitemOpen
  \bibfield  {author} {\bibinfo {author} {\bibfnamefont {A.}~\bibnamefont
  {Chacon}}, \bibinfo {author} {\bibfnamefont {L}~\bibnamefont {Heinen}},
  \bibinfo {author} {\bibfnamefont {M}~\bibnamefont {Halder}}, \bibinfo
  {author} {\bibfnamefont {A}~\bibnamefont {Bauer}}, \bibinfo {author}
  {\bibfnamefont {W}~\bibnamefont {Simeth}}, \bibinfo {author} {\bibfnamefont
  {S}~\bibnamefont {M{\"u}hlbauer}}, \bibinfo {author} {\bibfnamefont
  {H.}~\bibnamefont {Berger}}, \bibinfo {author} {\bibfnamefont {Markus}\
  \bibnamefont {Garst}}, \bibinfo {author} {\bibfnamefont {Achim}\ \bibnamefont
  {Rosch}}, \ and\ \bibinfo {author} {\bibfnamefont {Christian}\ \bibnamefont
  {Pfleiderer}},\ }\bibfield  {title} {\enquote {\bibinfo {title} {{Observation
  of two independent skyrmion phases in a chiral magnetic material}},}\
  }\href@noop {} {\bibfield  {journal} {\bibinfo  {journal} {Nature Physics}\
  }\textbf {\bibinfo {volume} {31}},\ \bibinfo {pages} {1} (\bibinfo {year}
  {2018})}\BibitemShut {NoStop}%
\bibitem [{\citenamefont {Qian}\ \emph {et~al.}(2018)\citenamefont {Qian},
  \citenamefont {Bannenberg}, \citenamefont {Wilhelm}, \citenamefont
  {Chaboussant}, \citenamefont {Debeer-Schmitt}, \citenamefont {Schmidt},
  \citenamefont {Aqeel}, \citenamefont {Palstra}, \citenamefont {Br{\"u}ck},
  \citenamefont {Lefering}, \citenamefont {Pappas}, \citenamefont {Mostovoy},\
  and\ \citenamefont {Leonov}}]{Qianeaat7323}%
  \BibitemOpen
  \bibfield  {author} {\bibinfo {author} {\bibfnamefont {Fengjiao}\
  \bibnamefont {Qian}}, \bibinfo {author} {\bibfnamefont {Lars~J.}\
  \bibnamefont {Bannenberg}}, \bibinfo {author} {\bibfnamefont {Heribert}\
  \bibnamefont {Wilhelm}}, \bibinfo {author} {\bibfnamefont {Gr{\'e}gory}\
  \bibnamefont {Chaboussant}}, \bibinfo {author} {\bibfnamefont {Lisa~M.}\
  \bibnamefont {Debeer-Schmitt}}, \bibinfo {author} {\bibfnamefont {Marcus~P.}\
  \bibnamefont {Schmidt}}, \bibinfo {author} {\bibfnamefont {Aisha}\
  \bibnamefont {Aqeel}}, \bibinfo {author} {\bibfnamefont {Thomas T.~M.}\
  \bibnamefont {Palstra}}, \bibinfo {author} {\bibfnamefont {Ekkes}\
  \bibnamefont {Br{\"u}ck}}, \bibinfo {author} {\bibfnamefont {Anton J.~E.}\
  \bibnamefont {Lefering}}, \bibinfo {author} {\bibfnamefont {Catherine}\
  \bibnamefont {Pappas}}, \bibinfo {author} {\bibfnamefont {Maxim}\
  \bibnamefont {Mostovoy}}, \ and\ \bibinfo {author} {\bibfnamefont
  {Andrey~O.}\ \bibnamefont {Leonov}},\ }\bibfield  {title} {\enquote {\bibinfo
  {title} {New magnetic phase of the chiral skyrmion material
  {C}u$_{2}${OS}e{O}$_{3}$},}\ }\href {\doibase 10.1126/sciadv.aat7323}
  {\bibfield  {journal} {\bibinfo  {journal} {Science Advances}\ }\textbf
  {\bibinfo {volume} {4}} (\bibinfo {year} {2018}),\
  10.1126/sciadv.aat7323}\BibitemShut {NoStop}%
\bibitem [{\citenamefont {Han}\ \emph {et~al.}(2020)\citenamefont {Han},
  \citenamefont {Garlow}, \citenamefont {Kharkov}, \citenamefont {Camacho},
  \citenamefont {Rov}, \citenamefont {Sauceda}, \citenamefont {Vats},
  \citenamefont {Kisslinger}, \citenamefont {Kato}, \citenamefont {Sushkov},
  \citenamefont {Zhu}, \citenamefont {Ulrich}, \citenamefont {S{\"o}hnel},\
  and\ \citenamefont {Seidel}}]{Han_2020}%
  \BibitemOpen
  \bibfield  {author} {\bibinfo {author} {\bibfnamefont {M.-G.}\ \bibnamefont
  {Han}}, \bibinfo {author} {\bibfnamefont {J.~A.}\ \bibnamefont {Garlow}},
  \bibinfo {author} {\bibfnamefont {Y.}~\bibnamefont {Kharkov}}, \bibinfo
  {author} {\bibfnamefont {L.}~\bibnamefont {Camacho}}, \bibinfo {author}
  {\bibfnamefont {R.}~\bibnamefont {Rov}}, \bibinfo {author} {\bibfnamefont
  {J.}~\bibnamefont {Sauceda}}, \bibinfo {author} {\bibfnamefont
  {G.}~\bibnamefont {Vats}}, \bibinfo {author} {\bibfnamefont {K.}~\bibnamefont
  {Kisslinger}}, \bibinfo {author} {\bibfnamefont {T.}~\bibnamefont {Kato}},
  \bibinfo {author} {\bibfnamefont {O.}~\bibnamefont {Sushkov}}, \bibinfo
  {author} {\bibfnamefont {Y.}~\bibnamefont {Zhu}}, \bibinfo {author}
  {\bibfnamefont {C.}~\bibnamefont {Ulrich}}, \bibinfo {author} {\bibfnamefont
  {T.}~\bibnamefont {S{\"o}hnel}}, \ and\ \bibinfo {author} {\bibfnamefont
  {J.}~\bibnamefont {Seidel}},\ }\bibfield  {title} {\enquote {\bibinfo {title}
  {Scaling, rotation, and channeling behavior of helical and skyrmion spin
  textures in thin films of {T}e-doped {C}u$_{2}${OS}e{O}$_{3}$},}\ }\href
  {\doibase 10.1126/sciadv.aax2138} {\bibfield  {journal} {\bibinfo  {journal}
  {Science Advances}\ }\textbf {\bibinfo {volume} {6}} (\bibinfo {year}
  {2020}),\ 10.1126/sciadv.aax2138}\BibitemShut {NoStop}%
\bibitem [{\citenamefont {Milde}\ \emph {et~al.}(2016)\citenamefont {Milde},
  \citenamefont {Neuber}, \citenamefont {Bauer}, \citenamefont {Pfleiderer},
  \citenamefont {Berger},\ and\ \citenamefont {Eng}}]{Milde2016}%
  \BibitemOpen
  \bibfield  {author} {\bibinfo {author} {\bibfnamefont {P.}~\bibnamefont
  {Milde}}, \bibinfo {author} {\bibfnamefont {E.}~\bibnamefont {Neuber}},
  \bibinfo {author} {\bibfnamefont {A.}~\bibnamefont {Bauer}}, \bibinfo
  {author} {\bibfnamefont {C.}~\bibnamefont {Pfleiderer}}, \bibinfo {author}
  {\bibfnamefont {H.}~\bibnamefont {Berger}}, \ and\ \bibinfo {author}
  {\bibfnamefont {L.M.}\ \bibnamefont {Eng}},\ }\bibfield  {title} {\enquote
  {\bibinfo {title} {Heuristic description of magnetoelectricity of
  {C}u$_{2}${OS}e{O}$_{3}$},}\ }\href@noop {} {\bibfield  {journal} {\bibinfo
  {journal} {Nano Lett.}\ }\textbf {\bibinfo {volume} {16}},\ \bibinfo {pages}
  {5612} (\bibinfo {year} {2016})}\BibitemShut {NoStop}%
\bibitem [{\citenamefont {Wadas}\ \emph {et~al.}(1995)\citenamefont {Wadas},
  \citenamefont {Wiesendanger},\ and\ \citenamefont
  {Novotny}}]{1995:Wadas:JApplPhys}%
  \BibitemOpen
  \bibfield  {author} {\bibinfo {author} {\bibfnamefont {A.}~\bibnamefont
  {Wadas}}, \bibinfo {author} {\bibfnamefont {R.}~\bibnamefont {Wiesendanger}},
  \ and\ \bibinfo {author} {\bibfnamefont {P.}~\bibnamefont {Novotny}},\
  }\bibfield  {title} {\enquote {\bibinfo {title} {{Bubble domains in garnet
  films studied by magnetic force microscopy}},}\ }\href {\doibase
  10.1063/1.360513} {\bibfield  {journal} {\bibinfo  {journal} {J. Appl.
  Phys.}\ }\textbf {\bibinfo {volume} {78}},\ \bibinfo {pages} {6324} (\bibinfo
  {year} {1995})}\BibitemShut {NoStop}%
\bibitem [{\citenamefont {Milde}\ \emph {et~al.}(2013)\citenamefont {Milde},
  \citenamefont {K\"ohler}, \citenamefont {Seidel}, \citenamefont {Eng},
  \citenamefont {Bauer}, \citenamefont {Chacon}, \citenamefont {Kindervater},
  \citenamefont {M\"uhlbauer}, \citenamefont {Pfleiderer}, \citenamefont
  {Sch\"utte},\ and\ \citenamefont {Rosch}}]{Milde2013}%
  \BibitemOpen
  \bibfield  {author} {\bibinfo {author} {\bibfnamefont {P.}~\bibnamefont
  {Milde}}, \bibinfo {author} {\bibfnamefont {D.}~\bibnamefont {K\"ohler}},
  \bibinfo {author} {\bibfnamefont {J.}~\bibnamefont {Seidel}}, \bibinfo
  {author} {\bibfnamefont {L.M.}\ \bibnamefont {Eng}}, \bibinfo {author}
  {\bibfnamefont {A.}~\bibnamefont {Bauer}}, \bibinfo {author} {\bibfnamefont
  {A.}~\bibnamefont {Chacon}}, \bibinfo {author} {\bibfnamefont
  {J.}~\bibnamefont {Kindervater}}, \bibinfo {author} {\bibfnamefont
  {S.}~\bibnamefont {M\"uhlbauer}}, \bibinfo {author} {\bibfnamefont
  {C.}~\bibnamefont {Pfleiderer}}, \bibinfo {author} {\bibfnamefont
  {C.}~\bibnamefont {Sch\"utte}}, \ and\ \bibinfo {author} {\bibfnamefont
  {A.}~\bibnamefont {Rosch}},\ }\bibfield  {title} {\enquote {\bibinfo {title}
  {Unwinding of a skyrmion lattice by magnetic monopoles},}\ }\href@noop {}
  {\bibfield  {journal} {\bibinfo  {journal} {Science}\ }\textbf {\bibinfo
  {volume} {340}},\ \bibinfo {pages} {1076} (\bibinfo {year}
  {2013})}\BibitemShut {NoStop}%
\bibitem [{\citenamefont {K\'ezsm\'arki}\ \emph {et~al.}(2015)\citenamefont
  {K\'ezsm\'arki}, \citenamefont {Bord\'acs}, \citenamefont {Milde},
  \citenamefont {Neuber}, \citenamefont {Eng}, \citenamefont {White},
  \citenamefont {R\o{}nnow}, \citenamefont {Dewhurst}, \citenamefont
  {Mochizuki}, \citenamefont {Yanai}, \citenamefont {Nakamura}, \citenamefont
  {Ehlers}, \citenamefont {Tsurkan},\ and\ \citenamefont
  {Loidl}}]{Kezsmarki2015}%
  \BibitemOpen
  \bibfield  {author} {\bibinfo {author} {\bibfnamefont {I.}~\bibnamefont
  {K\'ezsm\'arki}}, \bibinfo {author} {\bibfnamefont {S.}~\bibnamefont
  {Bord\'acs}}, \bibinfo {author} {\bibfnamefont {P.}~\bibnamefont {Milde}},
  \bibinfo {author} {\bibfnamefont {E.}~\bibnamefont {Neuber}}, \bibinfo
  {author} {\bibfnamefont {L.M.}\ \bibnamefont {Eng}}, \bibinfo {author}
  {\bibfnamefont {J.S.}\ \bibnamefont {White}}, \bibinfo {author}
  {\bibfnamefont {H.M.}\ \bibnamefont {R\o{}nnow}}, \bibinfo {author}
  {\bibfnamefont {C.D.}\ \bibnamefont {Dewhurst}}, \bibinfo {author}
  {\bibfnamefont {M.}~\bibnamefont {Mochizuki}}, \bibinfo {author}
  {\bibfnamefont {K.}~\bibnamefont {Yanai}}, \bibinfo {author} {\bibfnamefont
  {H.}~\bibnamefont {Nakamura}}, \bibinfo {author} {\bibfnamefont
  {D.}~\bibnamefont {Ehlers}}, \bibinfo {author} {\bibfnamefont
  {V.}~\bibnamefont {Tsurkan}}, \ and\ \bibinfo {author} {\bibfnamefont
  {A.}~\bibnamefont {Loidl}},\ }\bibfield  {title} {\enquote {\bibinfo {title}
  {N\'el-type skyrmion lattice with confined orientation in the polar magnetic
  semiconductor gav$_{4}$s$_{8}$},}\ }\href@noop {} {\bibfield  {journal}
  {\bibinfo  {journal} {Nat. Mat.}\ }\textbf {\bibinfo {volume} {14}},\
  \bibinfo {pages} {1116} (\bibinfo {year} {2015})}\BibitemShut {NoStop}%
\bibitem [{\citenamefont {Zhang}\ \emph {et~al.}(2016)\citenamefont {Zhang},
  \citenamefont {Bauer}, \citenamefont {Burn}, \citenamefont {Milde},
  \citenamefont {Neuber}, \citenamefont {Eng}, \citenamefont {Berger},
  \citenamefont {Pfleiderer}, \citenamefont {van~der Laan},\ and\ \citenamefont
  {Hesjedal}}]{Zhang2016}%
  \BibitemOpen
  \bibfield  {author} {\bibinfo {author} {\bibfnamefont {S.~L.}\ \bibnamefont
  {Zhang}}, \bibinfo {author} {\bibfnamefont {A.}~\bibnamefont {Bauer}},
  \bibinfo {author} {\bibfnamefont {D.~M.}\ \bibnamefont {Burn}}, \bibinfo
  {author} {\bibfnamefont {P.}~\bibnamefont {Milde}}, \bibinfo {author}
  {\bibfnamefont {E.}~\bibnamefont {Neuber}}, \bibinfo {author} {\bibfnamefont
  {L.~M.}\ \bibnamefont {Eng}}, \bibinfo {author} {\bibfnamefont
  {H.}~\bibnamefont {Berger}}, \bibinfo {author} {\bibfnamefont
  {C.}~\bibnamefont {Pfleiderer}}, \bibinfo {author} {\bibfnamefont
  {G.}~\bibnamefont {van~der Laan}}, \ and\ \bibinfo {author} {\bibfnamefont
  {T.}~\bibnamefont {Hesjedal}},\ }\bibfield  {title} {\enquote {\bibinfo
  {title} {Multidomain skyrmion lattice state in {C}u$_{2}${OS}e{O}$_{3}$},}\
  }\href@noop {} {\bibfield  {journal} {\bibinfo  {journal} {Nano Letters}\
  }\textbf {\bibinfo {volume} {16}},\ \bibinfo {pages} {3285} (\bibinfo {year}
  {2016})}\BibitemShut {NoStop}%
\bibitem [{\citenamefont {Baćani}\ \emph {et~al.}(2016)\citenamefont
  {Baćani}, \citenamefont {Marioni}, \citenamefont {Schwenk},\ and\
  \citenamefont {Hug}}]{BacaniHug_2016}%
  \BibitemOpen
  \bibfield  {author} {\bibinfo {author} {\bibfnamefont {Mirko}\ \bibnamefont
  {Baćani}}, \bibinfo {author} {\bibfnamefont {Miguel}\ \bibnamefont
  {Marioni}}, \bibinfo {author} {\bibfnamefont {J.}~\bibnamefont {Schwenk}}, \
  and\ \bibinfo {author} {\bibfnamefont {Hans}\ \bibnamefont {Hug}},\
  }\bibfield  {title} {\enquote {\bibinfo {title} {How to measure the local
  {D}zyaloshinskii-{M}oriya {I}nteraction in {S}kyrmion {T}hin {F}ilm
  {M}ultilayers},}\ }\href {\doibase 10.1038/s41598-019-39501-x} {\bibfield
  {journal} {\bibinfo  {journal} {Scientific Reports}\ }\textbf {\bibinfo
  {volume} {9}} (\bibinfo {year} {2016}),\
  10.1038/s41598-019-39501-x}\BibitemShut {NoStop}%
\bibitem [{\citenamefont {Masapogu}\ \emph {et~al.}(2019)\citenamefont
  {Masapogu}, \citenamefont {Yagil}, \citenamefont {Soumyanarayanan},
  \citenamefont {K.C.~Tan}, \citenamefont {Almoalem}, \citenamefont {Ma},
  \citenamefont {Auslaender},\ and\ \citenamefont
  {Panagopoulos}}]{MasapoguPanagopoulos_2019}%
  \BibitemOpen
  \bibfield  {author} {\bibinfo {author} {\bibfnamefont {Raju}\ \bibnamefont
  {Masapogu}}, \bibinfo {author} {\bibfnamefont {Alon}\ \bibnamefont {Yagil}},
  \bibinfo {author} {\bibfnamefont {Anjan}\ \bibnamefont {Soumyanarayanan}},
  \bibinfo {author} {\bibfnamefont {Anthony}\ \bibnamefont {K.C.~Tan}},
  \bibinfo {author} {\bibfnamefont {Avior}\ \bibnamefont {Almoalem}}, \bibinfo
  {author} {\bibfnamefont {Fusheng}\ \bibnamefont {Ma}}, \bibinfo {author}
  {\bibfnamefont {Ophir}\ \bibnamefont {Auslaender}}, \ and\ \bibinfo {author}
  {\bibfnamefont {C.}~\bibnamefont {Panagopoulos}},\ }\bibfield  {title}
  {\enquote {\bibinfo {title} {The evolution of skyrmions in
  {I}r/{F}e/{C}o/{P}t multilayers and their topological hall signature},}\
  }\href {\doibase 10.1038/s41467-018-08041-9} {\bibfield  {journal} {\bibinfo
  {journal} {Nature Communications}\ }\textbf {\bibinfo {volume} {10}},\
  \bibinfo {pages} {696} (\bibinfo {year} {2019})}\BibitemShut {NoStop}%
\bibitem [{\citenamefont {Weaver}\ and\ \citenamefont
  {Abraham}(1991)}]{Weaver1991}%
  \BibitemOpen
  \bibfield  {author} {\bibinfo {author} {\bibfnamefont {M.R.}\ \bibnamefont
  {Weaver}}\ and\ \bibinfo {author} {\bibfnamefont {D.W.}\ \bibnamefont
  {Abraham}},\ }\bibfield  {title} {\enquote {\bibinfo {title} {{High
  resolution atomic force microscopy potentiometry}},}\ }\href@noop {}
  {\bibfield  {journal} {\bibinfo  {journal} {J. Vac. Sci. Technol. B}\
  }\textbf {\bibinfo {volume} {10598}},\ \bibinfo {pages} {1559} (\bibinfo
  {year} {1991})}\BibitemShut {NoStop}%
\bibitem [{\citenamefont {Nonnenmacher}\ \emph {et~al.}(1991)\citenamefont
  {Nonnenmacher}, \citenamefont {O’Boyle},\ and\ \citenamefont
  {Wickramasinghe}}]{Nonnenmacher1991}%
  \BibitemOpen
  \bibfield  {author} {\bibinfo {author} {\bibfnamefont {M.}~\bibnamefont
  {Nonnenmacher}}, \bibinfo {author} {\bibfnamefont {M.P.}\ \bibnamefont
  {O’Boyle}}, \ and\ \bibinfo {author} {\bibfnamefont {H.K.}\ \bibnamefont
  {Wickramasinghe}},\ }\bibfield  {title} {\enquote {\bibinfo {title} {{Kelvin
  probe force microscopy}},}\ }\href@noop {} {\bibfield  {journal} {\bibinfo
  {journal} {Appl. Phys. Lett.}\ }\textbf {\bibinfo {volume} {58}},\ \bibinfo
  {pages} {2921--2923} (\bibinfo {year} {1991})}\BibitemShut {NoStop}%
\bibitem [{\citenamefont {Zerweck}\ \emph {et~al.}(2005)\citenamefont
  {Zerweck}, \citenamefont {Loppacher}, \citenamefont {Otto}, \citenamefont
  {Grafstr{\"o}m},\ and\ \citenamefont {Eng}}]{Zerweck2005}%
  \BibitemOpen
  \bibfield  {author} {\bibinfo {author} {\bibfnamefont {U.}~\bibnamefont
  {Zerweck}}, \bibinfo {author} {\bibfnamefont {Ch.}\ \bibnamefont
  {Loppacher}}, \bibinfo {author} {\bibfnamefont {T.}~\bibnamefont {Otto}},
  \bibinfo {author} {\bibfnamefont {S.}~\bibnamefont {Grafstr{\"o}m}}, \ and\
  \bibinfo {author} {\bibfnamefont {L.M.}\ \bibnamefont {Eng}},\ }\bibfield
  {title} {\enquote {\bibinfo {title} {{A}ccuracy and resolution limits of
  {K}elvin probe force microscopy},}\ }\href@noop {} {\bibfield  {journal}
  {\bibinfo  {journal} {Phys. Rev. B}\ }\textbf {\bibinfo {volume} {71}},\
  \bibinfo {pages} {125424} (\bibinfo {year} {2005})}\BibitemShut {NoStop}%
\bibitem [{Omi()}]{Omicron}%
  \BibitemOpen
  \href@noop {} {\enquote {\bibinfo {title} {{O}micron {N}ano{T}echnology
  {G}mbh, {T}aunusstein, {G}ermany},}\ }\BibitemShut {NoStop}%
\bibitem [{RHK()}]{RHK}%
  \BibitemOpen
  \href@noop {} {\enquote {\bibinfo {title} {{RHK} {T}echnology, {I}nc., 1050
  {E}ast {M}aple {R}oad, {T}roy, {MI} 48083 {USA}},}\ }\BibitemShut {NoStop}%
\bibitem [{Nan()}]{Nanosensors}%
  \BibitemOpen
  \href@noop {} {\enquote {\bibinfo {title} {{NANOSENSORS}\texttrademark, {R}ue
  {J}aquet-{D}roz 1, {C}ase {P}ostale 216, {CH}-2002 {N}euchatel,
  {S}witzerland},}\ }\BibitemShut {NoStop}%
\bibitem [{\citenamefont {Schwarze}\ \emph {et~al.}(2015)\citenamefont
  {Schwarze}, \citenamefont {Waizner}, \citenamefont {Garst}, \citenamefont
  {Bauer}, \citenamefont {Stasinopoulos}, \citenamefont {Berger}, \citenamefont
  {Rosch}, \citenamefont {Pfleiderer},\ and\ \citenamefont
  {Grundler}}]{2015:Schwarze:NatureMater}%
  \BibitemOpen
  \bibfield  {author} {\bibinfo {author} {\bibfnamefont {T.}~\bibnamefont
  {Schwarze}}, \bibinfo {author} {\bibfnamefont {J.}~\bibnamefont {Waizner}},
  \bibinfo {author} {\bibfnamefont {M.}~\bibnamefont {Garst}}, \bibinfo
  {author} {\bibfnamefont {A.}~\bibnamefont {Bauer}}, \bibinfo {author}
  {\bibfnamefont {I.}~\bibnamefont {Stasinopoulos}}, \bibinfo {author}
  {\bibfnamefont {H.}~\bibnamefont {Berger}}, \bibinfo {author} {\bibfnamefont
  {A.}~\bibnamefont {Rosch}}, \bibinfo {author} {\bibfnamefont
  {C.}~\bibnamefont {Pfleiderer}}, \ and\ \bibinfo {author} {\bibfnamefont
  {D.}~\bibnamefont {Grundler}},\ }\bibfield  {title} {\enquote {\bibinfo
  {title} {{Universal helimagnon and skyrmion excitations in metallic,
  semiconducting and insulating chiral magnets}},}\ }\href {\doibase
  10.1038/nmat4223} {\bibfield  {journal} {\bibinfo  {journal} {Nature Mater.}\
  }\textbf {\bibinfo {volume} {14}},\ \bibinfo {pages} {478} (\bibinfo {year}
  {2015})}\BibitemShut {NoStop}%
\bibitem [{\citenamefont {Grigoriev}\ \emph {et~al.}(2006)\citenamefont
  {Grigoriev}, \citenamefont {Maleyev}, \citenamefont {Okorokov}, \citenamefont
  {Chetverikov}, \citenamefont {B\"oni}, \citenamefont {Georgii}, \citenamefont
  {Lamago}, \citenamefont {Eckerlebe},\ and\ \citenamefont
  {Pranzas}}]{Grigoriev_2006}%
  \BibitemOpen
  \bibfield  {author} {\bibinfo {author} {\bibfnamefont {S.~V.}\ \bibnamefont
  {Grigoriev}}, \bibinfo {author} {\bibfnamefont {S.~V.}\ \bibnamefont
  {Maleyev}}, \bibinfo {author} {\bibfnamefont {A.~I.}\ \bibnamefont
  {Okorokov}}, \bibinfo {author} {\bibfnamefont {Yu.~O.}\ \bibnamefont
  {Chetverikov}}, \bibinfo {author} {\bibfnamefont {P.}~\bibnamefont {B\"oni}},
  \bibinfo {author} {\bibfnamefont {R.}~\bibnamefont {Georgii}}, \bibinfo
  {author} {\bibfnamefont {D.}~\bibnamefont {Lamago}}, \bibinfo {author}
  {\bibfnamefont {H.}~\bibnamefont {Eckerlebe}}, \ and\ \bibinfo {author}
  {\bibfnamefont {K.}~\bibnamefont {Pranzas}},\ }\bibfield  {title} {\enquote
  {\bibinfo {title} {Magnetic structure of {M}n{S}i under an applied field
  probed by polarized small-angle neutron scattering},}\ }\href {\doibase
  10.1103/PhysRevB.74.214414} {\bibfield  {journal} {\bibinfo  {journal} {Phys.
  Rev. B}\ }\textbf {\bibinfo {volume} {74}},\ \bibinfo {pages} {214414}
  (\bibinfo {year} {2006})}\BibitemShut {NoStop}%
\bibitem [{\citenamefont {Kousaka}\ \emph {et~al.}(2014)\citenamefont
  {Kousaka}, \citenamefont {Ikeda}, \citenamefont {Ogura}, \citenamefont
  {Yoshii}, \citenamefont {Akimitsu}, \citenamefont {Ohishi}, \citenamefont
  {Suzuki}, \citenamefont {Hiraka}, \citenamefont {Miyagawa}, \citenamefont
  {Nishihara}, \citenamefont {Inoue},\ and\ \citenamefont
  {Kishine}}]{Kousaka_2014_article}%
  \BibitemOpen
  \bibfield  {author} {\bibinfo {author} {\bibfnamefont {Yusuke}\ \bibnamefont
  {Kousaka}}, \bibinfo {author} {\bibfnamefont {Naoki}\ \bibnamefont {Ikeda}},
  \bibinfo {author} {\bibfnamefont {Takahiro}\ \bibnamefont {Ogura}}, \bibinfo
  {author} {\bibfnamefont {Toha}\ \bibnamefont {Yoshii}}, \bibinfo {author}
  {\bibfnamefont {Jun}\ \bibnamefont {Akimitsu}}, \bibinfo {author}
  {\bibfnamefont {Kazuki}\ \bibnamefont {Ohishi}}, \bibinfo {author}
  {\bibfnamefont {Junichi}\ \bibnamefont {Suzuki}}, \bibinfo {author}
  {\bibfnamefont {Haruhiko}\ \bibnamefont {Hiraka}}, \bibinfo {author}
  {\bibfnamefont {Marina}\ \bibnamefont {Miyagawa}}, \bibinfo {author}
  {\bibfnamefont {Sadafumi}\ \bibnamefont {Nishihara}}, \bibinfo {author}
  {\bibfnamefont {Katsuya}\ \bibnamefont {Inoue}}, \ and\ \bibinfo {author}
  {\bibfnamefont {Junichiro}\ \bibnamefont {Kishine}},\ }\bibfield  {title}
  {\enquote {\bibinfo {title} {Chiral {M}agnetic {S}oliton {L}attice in
  {M}n{S}i},}\ }\href {\doibase 10.7566/JPSCP.2.010205} {\bibfield  {journal}
  {\bibinfo  {journal} {JPS Conf. Proc.}\ }\textbf {\bibinfo {volume} {2}},\
  \bibinfo {pages} {010205} (\bibinfo {year} {2014})}\BibitemShut {NoStop}%
\bibitem [{\citenamefont {Aqeel}\ \emph {et~al.}(2016)\citenamefont {Aqeel},
  \citenamefont {Vlietstra}, \citenamefont {Roy}, \citenamefont {Mostovoy},
  \citenamefont {van Wees},\ and\ \citenamefont {Palstra}}]{Aqeel2016}%
  \BibitemOpen
  \bibfield  {author} {\bibinfo {author} {\bibfnamefont {A.}~\bibnamefont
  {Aqeel}}, \bibinfo {author} {\bibfnamefont {N.}~\bibnamefont {Vlietstra}},
  \bibinfo {author} {\bibfnamefont {A.}~\bibnamefont {Roy}}, \bibinfo {author}
  {\bibfnamefont {M.}~\bibnamefont {Mostovoy}}, \bibinfo {author}
  {\bibfnamefont {B.~J.}\ \bibnamefont {van Wees}}, \ and\ \bibinfo {author}
  {\bibfnamefont {T.~T.~M.}\ \bibnamefont {Palstra}},\ }\bibfield  {title}
  {\enquote {\bibinfo {title} {Electrical detection of spiral spin structures
  in {P}t$|${C}u$_{2}${OS}e{O}$_{3}$ heterostructures},}\ }\href {\doibase
  10.1103/PhysRevB.94.134418} {\bibfield  {journal} {\bibinfo  {journal} {Phys.
  Rev. B}\ }\textbf {\bibinfo {volume} {94}},\ \bibinfo {pages} {134418}
  (\bibinfo {year} {2016})}\BibitemShut {NoStop}%
\bibitem [{\citenamefont {Bos}\ \emph {et~al.}(2008)\citenamefont {Bos},
  \citenamefont {Colin},\ and\ \citenamefont {Palstra}}]{Bos2008}%
  \BibitemOpen
  \bibfield  {author} {\bibinfo {author} {\bibfnamefont {J.-W.G.}\ \bibnamefont
  {Bos}}, \bibinfo {author} {\bibfnamefont {C.V.}\ \bibnamefont {Colin}}, \
  and\ \bibinfo {author} {\bibfnamefont {T.T.M.}\ \bibnamefont {Palstra}},\
  }\bibfield  {title} {\enquote {\bibinfo {title} {Magnetoelectric coupling in
  the cubic ferrimagnet {C}u$_{2}${OS}e{O}$_{3}$},}\ }\href@noop {} {\bibfield
  {journal} {\bibinfo  {journal} {Phys. Rev. B}\ }\textbf {\bibinfo {volume}
  {78}},\ \bibinfo {pages} {094416} (\bibinfo {year} {2008})}\BibitemShut
  {NoStop}%
\bibitem [{\citenamefont {Rybakov}\ \emph {et~al.}(2016)\citenamefont
  {Rybakov}, \citenamefont {Borisov}, \citenamefont {Blügel},\ and\
  \citenamefont {Kiselev}}]{Rybakov_2016}%
  \BibitemOpen
  \bibfield  {author} {\bibinfo {author} {\bibfnamefont {Filipp~N}\
  \bibnamefont {Rybakov}}, \bibinfo {author} {\bibfnamefont {Aleksandr~B}\
  \bibnamefont {Borisov}}, \bibinfo {author} {\bibfnamefont {Stefan}\
  \bibnamefont {Blügel}}, \ and\ \bibinfo {author} {\bibfnamefont {Nikolai~S}\
  \bibnamefont {Kiselev}},\ }\bibfield  {title} {\enquote {\bibinfo {title}
  {New spiral state and skyrmion lattice in 3{D} model of chiral magnets},}\
  }\href {\doibase 10.1088/1367-2630/18/4/045002} {\bibfield  {journal}
  {\bibinfo  {journal} {New Journal of Physics}\ }\textbf {\bibinfo {volume}
  {18}},\ \bibinfo {pages} {045002} (\bibinfo {year} {2016})}\BibitemShut
  {NoStop}%
\end{thebibliography}%
\end{document}